\documentclass[useAMS,usenatbib]{mn2e}
\usepackage{times}
\usepackage{graphicx}

\newcommand{\xmm}{{\it XMM~\/}}
\newcommand{\xmmn}{{\it XMM-Newton~\/}}

\newcommand{\chandra}{{\it Chandra~\/}}

\newcommand{\rosat}{{\it ROSAT~\/}}

\newcommand{\spitzer}{{\it Spitzer~\/}}
\newcommand{\aap}{{\it A\&A~\/}}

\newcommand{\apj}{{\it ApJ~\/}}

\newcommand{\mnras}{{\it MNRAS~\/}}

\def\Msun{\hbox{$\rm M_{\odot}~$}}
\def\ergcms{{\rm ~erg~cm^{-2}~s^{-1}}}

\def\ergsec{{\rm ~erg~s^{-1}}}
\def\kmsec{{\rm ~km~s^{-1}}}
\def\chisq{{$\chi^{2}$}}

\def\dg{^{\circ}}
\def\H0{{\rm ~km~s^{-1}~Mpc^{-1}}}

\def\eg{{\it e.g.,~\/}}

\def\ie{{\it i.e.,~\/}}
\def\cf{{\it c.f.~\/}}
\def\la{\mathrel{\hbox{\rlap{\hbox{\lower4pt\hbox{$\sim$}}}{\raise2pt\hbox{$<$}}
}}}
\def\ga{\mathrel{\hbox{\rlap{\hbox{\lower4pt\hbox{$\sim$}}}{\raise2pt\hbox{$>$}}
}}}
\def\d25{D$_{25}$}
\def\nh{{$N_{H}$}$~$}

\def\hii{H{\small II}$~$}

\def\deg{\hbox{$^\circ$~\/}}
\def\arcmin{\hbox{$^\prime$~\/}}
\def\arcsec{\hbox{$^{\prime\prime}$~\/}}

\begin{document}

\title[The X-ray properties of NGC 55 ] {The X-ray properties of the dwarf Magellanic-type galaxy NGC 55}
\author[A-M. Stobbart et al.]
    {A-M.\ Stobbart, T.P.\ Roberts, R.S.\ Warwick\\
X-ray \& Observational Astronomy Group, Dept. of Physics \& Astronomy,
University of Leicester, Leicester LE1 7RH, U.K.\\}

\date{Submitted}

\pagerange{\pageref{firstpage}--\pageref{lastpage}}
\pubyear{2004}

\maketitle

\label{firstpage}

\begin{abstract}

We present an analysis of the X-ray properties of the Magellanic-type
galaxy NGC 55 based on two contiguous \xmmn observations.  We detect a
total of 137 X-ray sources in the field of view, down to a flux of
$\sim 5 \times 10^{-15} \ergcms$ (0.3 -- 6 keV), 42 of which are
located within the optical confines of the galaxy.  On the basis of
X-ray colour classification and after correcting for background
objects, we conclude that our source sample includes $\sim 20$ X-ray
binaries, 5 supernova remnants and 7 very soft sources (including 2
good candidate supersoft sources) associated with NGC 55.  We also
detect an X-ray source coincident with a previously identified
globular cluster in NGC 55.  Detailed spectral and timing analysis was
carried out on 4 of the brightest X-ray sources (excluding the
brightest source, which was the subject of a previous paper). One of
these objects is identified with a Galactic foreground star and is a
possible new RS CVn system. The other three are persistent X-ray
sources with X-ray spectra well described by either a single absorbed
power-law ($\Gamma
\sim 2$) or a multicolour disc blackbody ($kT_{in} \sim 1$ keV) model.
While the observed luminosities of these sources ($L_X \sim 1 - 2
\times 10^{38} \ergsec$) and their X-ray spectra are consistent with
accreting X-ray binaries, further evidence of short term variability
is required to confirm this.  Although the {\it observed} X-ray
emission from NGC 55 is dominated by point sources, we do find
evidence of an underlying component, which is concentrated on the bar
region but has an extent of at least 6\arcmin (3 kpc) in the plane of
the galaxy and $\pm$1\arcmin ($\pm 500$ pc) perpendicular to it. This
emission is best fitted by a thermal plasma ({\sc mekal}) ($kT \sim
0.2$ keV) plus power-law ($\Gamma \sim 2$) model but with high
intrinsic absorption consistent with its location in the central disc
of the galaxy. We interpret the soft component as diffuse thermal
emission linked to regions of current star formation, whilst the hard
power-law component may originate in unresolved X-ray binary
sources. The {\it intrinsic} luminosity of this residual disc emission
may exceed $L_X \sim 6 \times 10^{38} \ergsec$ (0.3--6 keV).  A
comparison with other Magellanic systems confirms that, in terms of
both its discrete X-ray source population and its extended emission,
NGC 55 has X-ray properties which are typical of its class.

\end{abstract}


\section{Introduction}

NGC 55 is a member of the nearby Sculptor Group of galaxies, of which
the other prominent members are the starburst galaxy NGC 253 and the
spirals NGC 45, NGC 247, NGC 300 and NGC 7793.  NGC 55 itself has been
classified as a SB(s)m galaxy (\citealt{deVaucouleurs61}) and is
viewed almost edge-on with an optical extent of $32.4\arcmin
\times~5.6\arcmin$ (3rd Reference Catalog of Galaxies;
\citealt{deVaucouleurs91}).  The inclination estimates range from
$80\dg$ (\citealt*{hummel86}) to $85\dg$ (\citealt*{deVaucouleurs72});
here we assume a value of $81\dg$ (\citealt*{kisz88}). The optical
morphology of NGC 55 is rather asymmetric, with the brightest region
displaced $\sim 3\arcmin$ from the geometrical centre of the galaxy
(\citealt{robinson66}).  This feature has been interpreted as a bar
viewed near to end-on (\citealt{deVaucouleurs61}).  Although there is
active star formation present throughout much of the disc of NGC 55,
the radio continuum emission is concentrated on the bar region and at
6 cm is dominated by a triple source (e.g. \citealt{hummel86}). Recent
\spitzer far-infrared imaging suggests this is a young ($< 2$ Myr)
star formation complex (\citealt*{engelbracht04}).  The \spitzer
observations further imply a global star formation rate of 0.22
$M_{\odot}$ yr$^{-1}$ for NGC 55 (\citealt{engelbracht04}). The
edge-on orientation of NGC 55 affords us a prime view of the effects
of this disc-based star formation activity on the extra-planar
regions.  Spectacular ionized gas features, including giant shells and
possible galactic chimneys, protrude well above the plane of the
galaxy (\citealt*{ferguson96}; \citealt*{otte99}), suggesting that the
star formation powers the ejection of gas from the disc into the halo.
At least some of this gas appears to cool sufficiently to form new
stars in the halo (\citealt*{tullmann03}; \citealt*{tullmann04}).

For such a relatively nearby ($d = 1.78$ Mpc \footnote{The distance to
NGC 55 was estimated to be 1.45$^{+0.35}_{-0.30}$ Mpc
by \citet{graham82} based on the apparent magnitude of stars at the tip
of the red giant branch (the TRGB method), whereas \citet{pritchet87}
obtained a value of 1.34$\pm0.08$ Mpc, via photometry of carbon stars.
More recent estimates, using a variety of techniques, place NGC 55 at
a distance of between $1.7 - 2.1$ Mpc (\citealt{kara03};
\citealt*{tikhonov04};\citealt{steene04}).  In the present paper we
have taken the distance to NGC 55 to be 1.78 Mpc
(\citealt{kara03}).}), active star-forming system, NGC 55 has been
relatively poorly studied at X-ray wavelengths.  The first detailed
X-ray information came from \rosat PSPC and HRI observations
(\citealt*{read97}; \citealt{roberts97}; \citealt*{schlegel97};
\citealt*{dahlem98}), revealing a total of 25 sources in and around
the galaxy and evidence of localised diffuse emission. Of these
sources, 15 were located within the optical confines of the galaxy as
defined by the \d25 ellipse \citep{roberts97} with one particular
source, situated $\sim$7\arcmin to the east of the main bar complex,
found to be several times brighter than any other X-ray source
associated with the galaxy. New \xmmn observations establish this
object (XMMU J001528.9-391319) to be an ultraluminous X-ray source
(ULX) and, most probably, a black-hole X-ray binary system
(\citealt{stobbart04} - hereafter Paper I). Previous \chandra (ACIS-I)
observations detect possible extended X-ray emission above the disc of
the galaxy, as well as a somewhat brighter diffuse component
associated with the disc itself (\citealt{oshima02}).

Here we present a detailed analysis of the \xmmn observations of NGC
55, focusing on both the properties of the discrete X-ray source
population and the underlying diffuse emission.  The paper is set out
as follows.  We first describe the observations and preliminary data
reduction techniques (sec. 2). In sec. 3 we give details of the source
detection procedure and present a catalogue of the X-ray sources
detected in the NGC 55 field. The next three sections consider the
X-ray properties of four of the brightest X-ray sources in NGC 55
(sec. 4), the X-ray colours of the full set of detected sources
(sec. 5) and the morphology and spectral properties of the faint
diffuse X-ray emission observed in NGC 55 (sec. 6). We then discuss
the nature of the brightest X-ray sources in NGC 55 and consider the
overall X-ray properties of this galaxy in the context of other nearby
Magellanic-type systems (sec. 7). Finally, we briefly summarise our
results (sec. 8).

\section{Observations and data screening}

Two observations of NGC 55 were carried out by \xmmn in the period
2001 November 14-15, with the second observation commencing 2.2 ks
after the termination of the first. The pointings were offset in
opposite directions with respect to the centre of the galaxy so as to
image the full extent of the edge-on galaxy disc (see Table
\ref{table1}).

This paper focuses on data from the EPIC MOS and pn cameras
(\citealt{turner01}; \citealt{struder01}) taken in the full window
mode with the thin filter deployed. All the datasets were pipeline
processed and reduced using standard tools of the \xmmn Science
Analysis Software ({\sc sas}) version 5.4.1. After the removal of an
interval of soft proton flaring towards the end of the second pointing
and utilising only periods when all three cameras (MOS-1, MOS-2 and
pn) were in operation, the net exposure times were 30.4 ks and 21.5 ks
for the first and second observations respectively.

In the present analysis we use valid pn events with pattern 0-4 but
use pattern 0-12 for the MOS cameras. We also exclude events outside
the field of view and remove any hot pixels in the data by using the
flag expressions \#XMMEA\_EM and \#XMMEA\_EP for MOS and pn
respectively.  We also removed three additional hot columns which
remained in the pn data (two in the first observation, and one in the
second).

\begin{table}
\begin{center}
\caption{\label{table1}The \xmmn observations of NGC 55}
\begin{tabular}{ccccc}
\hline
Obs ID  &RA$^a$     &Dec$^a$        &Date & UT$_{start}$\\
\hline
0028740201  &00~15~46.0 &-39~15~28  &2001-11-14&14:20:08\\
0028740101  &00~14~32.9 &-39~10~46  &2001-11-15&00:24:43\\
\hline
\end{tabular}
\end{center}
\begin{tabular}{c}
$^a$ Epoch J2000 co-ordinates.
\end{tabular}
\end{table}

\section{X-ray Sources in the Field of NGC 55}

\subsection{Source detection}

The first requirement was to produce a list of the discrete X-ray
sources detected within the sky area covered by the two
\xmmn observations.  For this purpose we combined the MOS-1 and
MOS-2 events using the {\sc sas} task {\sc merge} and subsequently
used the combined MOS (\ie MOS-1 plus MOS-2) and pn data as separate
channels in the source detection process.  MOS and pn images were
produced for each observation in three energy bands, chosen to provide
a good signal-to-noise coverage of the EPIC data: 0.3--1 keV (soft),
1--2 keV (medium) and 2--6 keV (hard), together with the corresponding
exposure maps.  We believe that these energy bands will provide good
signal to noise detections for the sources in the field. Here we use a
pixel size of 4\arcsec $\times$ 4\arcsec.

The source detection routines were carried out separately for the MOS
and pn datasets over the three energy bands.  The {\sc sas} task {\sc
eboxdetect} was used to perform the initial source detection and
employs a sliding box detection method. The task has two different
modes of operation, the first of which is the local detection mode in
which the images are scanned by a sliding square box, and an object
centred in the box is classified as a source if its signal to noise is
greater than a specified threshold value\footnote{The signal is
derived from the pixel values inside a 5 $\times$ 5 pixel window, and
the local background and noise level is estimated from the surrounding
56 pixels (within a full 9 $\times$ 9 pixel window).}. Next, the
derived source positions were provided as input for the task {\sc
esplinemap} which constructs background maps using source-free regions
of the image. Once the background maps are available, they are
utilised in a further run of {\sc eboxdetect} in `map' mode, leading
to an improved detection sensitivity.  To complement the sliding-box
detection approach we also employed the wavelet search technique {via
\sc ewavelet}, a method which provides, amongst other advantages, good
suppression of multiple false detections in extended sources and is
also better at distinguishing close sources.  While for {\sc
eboxdetect} the source detection was performed simultaneously on the
soft, medium and hard datasets for a particular camera and
observation, the {\sc ewavelet} routine was performed separately on
each energy band, detector and observation.  Any additional sources
detected via the {\sc ewavelet} method were added to the {\sc
eboxdetect} list. As a final quality check we removed any sources we
deemed to be unreliable based on a visual inspection of the image (\eg
unconvincing sources at chip edges or gaps).

The next step was to use the task {\sc emldetect} to carry out, on a
source by source basis, a maximum likelihood fit of the point spread
function to the observed spatial distribution of counts measured in
the separate energy bands.  In this context accurate modelling of the
background is essential when determining the significance of a
detected source and at this point we used adaptively-smoothed
background maps obtained from source-excluded images via the {\sc
asmooth} task.  We ran the {\sc emldetect} task using these improved
background maps to produce a standardised set of source
parameters. Sources were classified as significant detections if they
had a likelihood (DET\_ML) value of 10 or more (corresponding to a
detection significance of $> 4 \sigma$) in at least one of the energy
bands, not just the summary band. The final source list was obtained
by completing two further iterations of the {\sc asmooth} and {\sc
emldetect} procedures.

Application of the above procedures resulted in four separate
sourcelists (\ie MOS and pn lists for each observation). We next
checked for astrometric offsets by correlating these lists with the
USNO A2.0 optical catalogue, using the task {\sc eposcorr}.  The
derived `optimum offsets' for each observation were used to correct
the X-ray source positions to the USNO frame of reference. The final
step was to merge the four {\sc emldetect} source lists into one
summary list using the {\sc srcmatch} task, which correlates sources
co-located within their 5$\sigma$ centroiding errors plus a 1\arcsec
systematic.

\subsection{The source catalogue}

\begin{figure*}
\begin{center}
\scalebox{0.47}{{\includegraphics[angle=0]{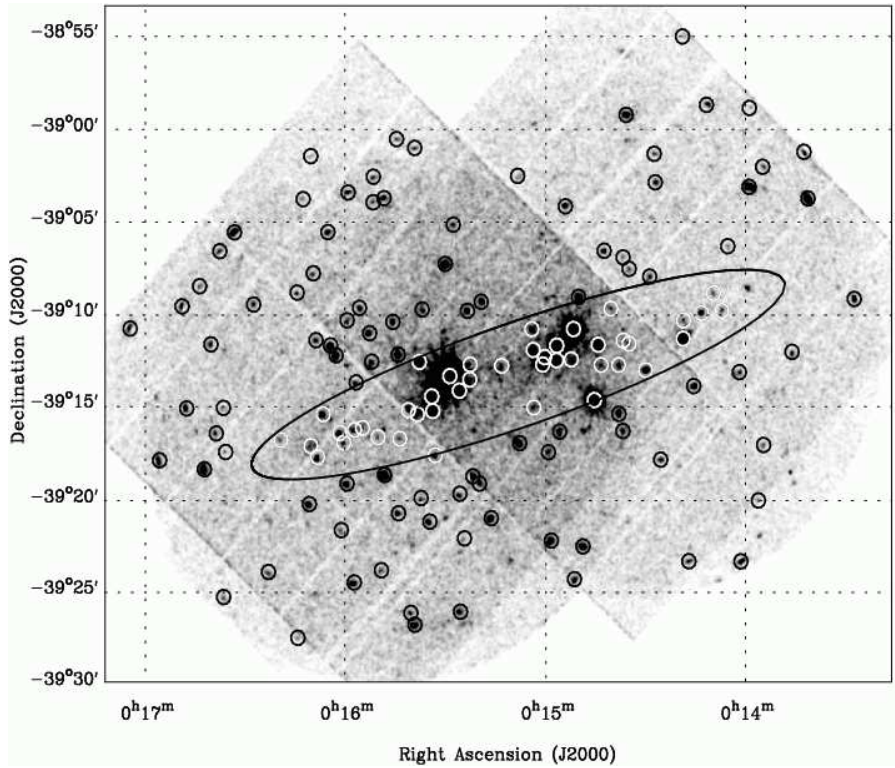}}}
\end{center}
\caption{\label{xbb} The \xmmn image of the NGC 55 field in a
broad (0.3--6 \,keV) bandpass.  The image has been lightly
smoothed using a circular Gaussian mask with $\sigma$=1 pixel
({\it i.e.,} 4$^{\prime\prime}$). Detected sources are identified
with circles. The optical \d25 ellipse of NGC 55 is also shown.}
\end{figure*}

\begin{figure*}
\begin{center}
\scalebox{0.47}{{\includegraphics[angle=0]{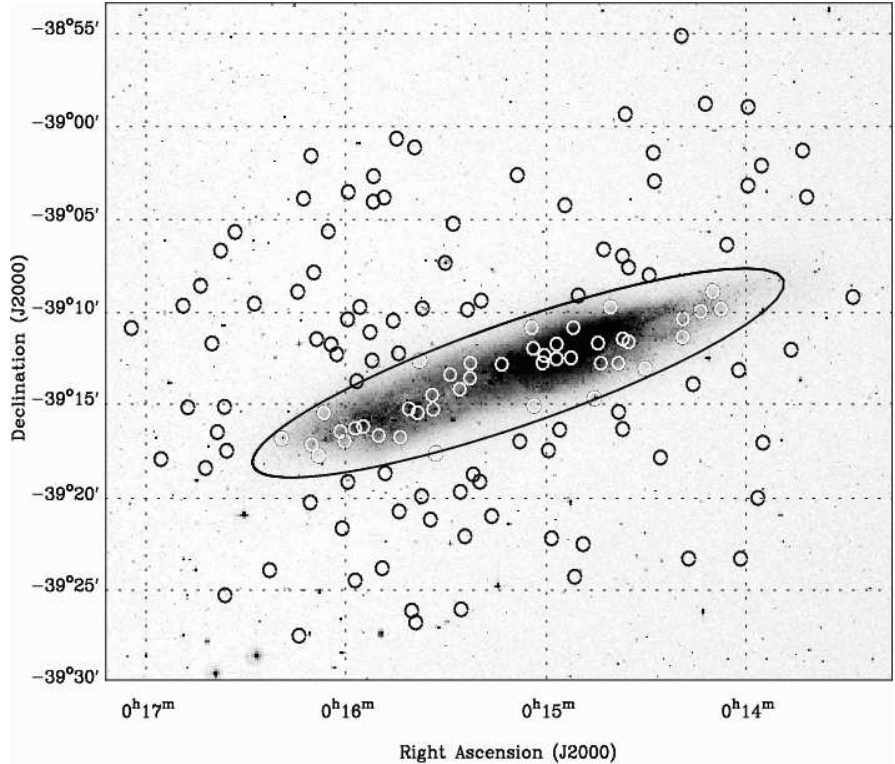}}}
\end{center}
\caption{\label{opt} The optical DSS-2 (blue) image of the NGC 55 field with
the X-ray source positions marked. The \d25 ellipse of NGC 55
is also shown.}
\end{figure*}

In total, 137 X-ray sources are detected within the field of view
encompassed by the two \xmmn observations, 42 of which are located
within the \d25 ellipse of NGC 55.  Fig. \ref{xbb} shows the broad
band (0.3--6 keV) X-ray image obtained by merging the MOS and pn data
from the two observations.  The positions of the 137 sources which
pass the significance test described earlier are marked with
circles. Fig. \ref{opt} shows the same X-ray source positions overlaid
on the optical DSS-2 blue image of the galaxy.


\begin{table*}
\begin{center}
\caption{\label{allsrcs} The full catalogue of sources detected in the \xmmn
observations}
{\scriptsize
\begin{tabular}{lcccccccccccccc}
\hline

SRC &RA		&DEC    	&r$_1\sigma$	&\multicolumn{3}{c}{pn cts $\rm ks^{-1}$}          	 &&\multicolumn{3}{c}{MOS cts $\rm ks^{-1}$}         &HR1        &HR2        &{\it f}$_X$        &L$_X$  \\ \cline{5-7} \cline{9-11}
ID  &(hh:mm:ss)  &(\deg:\arcmin:\arcsec)&(\arcsec)&S		 &M		    &H          	 &&S         &M          &H          &       &       &($\times 10^{-14}$)    &($\times 10^{36}$)\\

\hline

1   &00:13:28.37	&-39:09:03.0   &1.14   &--         	 &--         	    &--         	 &&\bf{3.4$\pm$0.4}  &\bf{3.0$\pm$0.4}   &\bf{1.4$\pm$0.4}   &-0.06$\pm$0.09 &-0.36$\pm$0.12 &4.63$\pm$0.54	&\\
2   &00:13:42.33	&-39:03:38.3   &0.60   &\bf{40.5$\pm$2.8}  &\bf{16.9$\pm$1.8}  &\bf{14.4$\pm$1.7}  &&\bf{13.2$\pm$1.1} &\bf{10.6$\pm$1.0}  &\bf{7.9$\pm$0.8}   &-0.28$\pm$0.05 &-0.12$\pm$0.08 &17.1$\pm$0.8   	&\\
3   &00:13:43.61	&-39:01:07.2   &0.87   &\bf{17.7$\pm$2.9}  &\bf{7.4$\pm$1.8}   &\bf{4.0$\pm$1.5}   &&\bf{4.2$\pm$0.6}  &\bf{2.2$\pm$0.5}   &0.0$\pm$0.1        &-0.36$\pm$0.12 &-0.88$\pm$0.15 &3.01$\pm$0.32	&\\
4   &00:13:46.83	&-39:11:55.7   &1.49   &\bf{4.3$\pm$0.8}   &\bf{2.3$\pm$0.5}   &\bf{1.3$\pm$0.4}   &&--            &--         &--         &-0.31$\pm$0.13 &-0.26$\pm$0.19 &1.51$\pm$0.27      			&\\
5   &00:13:55.16	&-39:16:58.4   &1.41   &\bf{2.1$\pm$0.6}   &\bf{1.2$\pm$0.3}   &\bf{2.0$\pm$0.5}   &&0.2$\pm$0.2       &\bf{0.9$\pm$0.2}   &0.4$\pm$0.2        &-0.01$\pm$0.19 &0.06 $\pm$0.23 &1.26$\pm$0.21	&\\
6   &00:13:55.87	&-39:01:56.5   &1.28   &\bf{2.6$\pm$0.8}   &\bf{1.9$\pm$0.6}   &1.0$\pm$0.6        &&\bf{1.3$\pm$0.3}  &\bf{1.1$\pm$0.3}   &0.9$\pm$0.3        &-0.12$\pm$0.19 &-0.17$\pm$0.27 &1.51$\pm$0.27	&\\
7   &00:13:56.62	&-39:19:58.3   &1.36   &0.5$\pm$0.4        &1.8$\pm$0.6        &\bf{2.9$\pm$0.8}   &&0.0$\pm$0.2       &0.8$\pm$0.2        &\bf{1.3$\pm$0.3}   &0.73$\pm$0.34  &0.23 $\pm$0.18 &2.01$\pm$0.30    	&\\
8   &00:13:59.88	&-39:03:01.8   &0.69   &\bf{28.7$\pm$2.0}  &\bf{11.3$\pm$1.2}  &\bf{6.3$\pm$1.0}   &&--            &--         &--         &-0.44$\pm$0.05 &-0.28$\pm$0.09 &8.13$\pm$0.60      			&\\
9   &00:13:59.90	&-38:58:46.0   &1.73   &\bf{4.5$\pm$1.0}   &0.0$\pm$0.3        &0.0$\pm$0.2        &&--            &--         &--         &-1.00$\pm$0.13 &--             &0.45$\pm$0.17      			&\\
10  &00:14:01.76	&-39:23:16.0   &0.72   &--         &--         &--             &&\bf{11.8$\pm$1.0} &\bf{8.4$\pm$0.9}   &\bf{4.5$\pm$0.6}   &-0.17$\pm$0.07 &-0.31$\pm$0.08 &14.6$\pm$1.0   			&\\
11  &00:14:02.49	&-39:13:02.5   &1.01   &\bf{3.8$\pm$0.6}   &\bf{1.4$\pm$0.3}   &0.1$\pm$0.2        &&\bf{0.7$\pm$0.2}  &0.4$\pm$0.1        &0.3$\pm$0.1        &-0.42$\pm$0.20 &-0.69$\pm$0.24 &0.68$\pm$0.10	&\\
12  &00:14:06.13	&-39:06:14.7   &1.91   &0.3$\pm$0.3        &\bf{1.2$\pm$0.3}   &0.8$\pm$0.4        &&--            &--         &--         &0.61$\pm$0.32  &-0.20$\pm$0.26 &0.68$\pm$0.22      			&\\
13$^*$  &00:14:07.84	&-39:09:44.0   &1.24   &\bf{0.8$\pm$0.3}   &\bf{1.4$\pm$0.4}   &\bf{1.2$\pm$0.3}   &&0.4$\pm$0.2       &\bf{0.7$\pm$0.2}   &0.4$\pm$0.2        &0.30$\pm$0.21  &-0.17$\pm$0.20 &0.99$\pm$0.15     &3.76\\
14$^*$  &00:14:10.33	&-39:08:47.3   &1.53   &0.7$\pm$0.4        &\bf{2.2$\pm$0.4}   &0.8$\pm$0.3        &&0.4$\pm$0.1       &\bf{0.8$\pm$0.2}   &0.3$\pm$0.2        &0.40$\pm$0.20  &-0.44$\pm$0.20 &0.88$\pm$0.15     &3.32\\
15  &00:14:12.66	&-38:58:36.6   &0.96   &\bf{8.3$\pm$1.2}   &\bf{3.7$\pm$0.8}   &2.4$\pm$0.7        &&\bf{2.7$\pm$0.6}  &\bf{2.0$\pm$0.5}   &\bf{2.2$\pm$0.6}   &-0.30$\pm$0.16 &-0.08$\pm$0.17 &3.24$\pm$0.38	&\\
16$^*$  &00:14:13.87	&-39:09:51.9   &0.83  &\bf{1.4$\pm$0.4}   &\bf{1.9$\pm$0.4}   &\bf{2.2$\pm$0.4}   &&0.5$\pm$0.1       &\bf{1.9$\pm$0.2}   &\bf{1.4$\pm$0.2}   &0.47$\pm$0.14  &-0.06$\pm$0.11 &2.07$\pm$0.20      &7.85\\
17  &00:14:16.12	&-39:13:48.5   &0.73   &\bf{5.8$\pm$0.7}   &\bf{2.8$\pm$0.4}   &\bf{1.8$\pm$0.4}   &&\bf{1.8$\pm$0.2}  &\bf{1.4$\pm$0.2}   &\bf{1.5$\pm$0.2}   &-0.26$\pm$0.11 &-0.09$\pm$0.12 &2.47$\pm$0.19    	&\\
18  &00:14:17.33	&-39:23:16.4   &1.16   &\bf{3.3$\pm$0.9}   &\bf{3.8$\pm$0.8}   &\bf{1.1$\pm$0.3}   &&\bf{2.6$\pm$0.5}  &\bf{1.5$\pm$0.4}   &0.5$\pm$0.3        &-0.13$\pm$0.15 &-0.53$\pm$0.19 &1.75$\pm$0.21     &\\
19$^*$  &00:14:19.24	&-39:10:16.2   &1.22   &0.0$\pm$0.2        &\bf{1.4$\pm$0.3}   &\bf{1.1$\pm$0.3}   &&0.3$\pm$0.1       &0.4$\pm$0.1        &\bf{0.7$\pm$0.2}   &0.65$\pm$0.26  &0.06 $\pm$0.18 &0.99$\pm$0.15     &3.76\\
20$^*$  &00:14:19.31	&-39:11:15.2   &0.53   &\bf{27.1$\pm$1.4}  &\bf{33.3$\pm$1.5}  &\bf{27.9$\pm$1.4}  &&\bf{8.1$\pm$0.5}  &\bf{18.4$\pm$0.8}  &\bf{16.5$\pm$0.8}  &0.24$\pm$0.03  &-0.07$\pm$0.03 &27.0$\pm$0.7      &102.4\\
21  &00:14:19.89	&-38:54:55.4   &2.46   &\bf{6.5$\pm$1.4}   &2.5$\pm$0.9        &1.2$\pm$0.9        &&--            &--         &--         &-0.44$\pm$0.17 &-0.35$\pm$0.35 &1.71$\pm$0.52      			&\\
22  &00:14:25.86	&-39:17:47.6   &1.12   &\bf{2.7$\pm$0.5}   &\bf{1.5$\pm$0.4}   &1.0$\pm$0.4        &&\bf{1.1$\pm$0.2}  &\bf{0.8$\pm$0.2}   &\bf{0.6$\pm$0.1}   &-0.23$\pm$0.15 &-0.19$\pm$0.19 &1.33$\pm$0.14	&\\
23  &00:14:27.81	&-39:02:48.6   &0.94   &\bf{3.7$\pm$0.7}   &\bf{1.6$\pm$0.4}   &\bf{1.6$\pm$0.4}   &&\bf{1.1$\pm$0.2}  &\bf{1.7$\pm$0.3}   &0.8$\pm$0.2        &-0.11$\pm$0.13 &-0.20$\pm$0.16 &1.79$\pm$0.21	&\\
24  &00:14:28.17	&-39:01:16.1   &1.46   &\bf{3.4$\pm$0.7}   &1.3$\pm$0.4        &0.4$\pm$0.3        &&\bf{0.9$\pm$0.2}  &0.6$\pm$0.2        &0.9$\pm$0.3        &-0.37$\pm$0.23 &-0.06$\pm$0.33 &0.98$\pm$0.19	&\\
25  &00:14:29.36	&-39:07:54.3   &1.23   &\bf{1.4$\pm$0.4}   &\bf{1.6$\pm$0.3}   &0.4$\pm$0.3        &&\bf{0.6$\pm$0.1}  &\bf{0.6$\pm$0.1}   &\bf{0.6$\pm$0.2}   &0.02$\pm$0.17  &-0.20$\pm$0.20 &0.87$\pm$0.14	&\\
26$^*$  &00:14:30.48	&-39:12:57.5   &0.69   &\bf{5.6$\pm$0.7}   &\bf{3.9$\pm$0.5}   &\bf{2.2$\pm$0.4}   &&\bf{1.3$\pm$0.2}  &\bf{2.3$\pm$0.3}   &\bf{1.1$\pm$0.2}   &0.03$\pm$0.09  &-0.31$\pm$0.10 &2.65$\pm$0.20  	&10.0\\
27$^*$  &00:14:35.45	&-39:11:32.2   &1.50   &\bf{2.2$\pm$0.5}   &0.0$\pm$0.2        &0.5$\pm$0.3        &&\bf{0.9$\pm$0.2}  &0.2$\pm$0.1        &0.0$\pm$0.1        &-0.85$\pm$0.18 &-0.19$\pm$0.69 &0.47$\pm$0.10     &1.78\\
28  &00:14:35.50	&-39:07:29.6   &1.66  &0.6$\pm$0.3        &0.6$\pm$0.3        &0.9$\pm$0.3        &&0.1$\pm$0.1       &\bf{0.7$\pm$0.2}   &0.5$\pm$0.1        &0.53$\pm$0.30  &-0.06$\pm$0.21 &0.76$\pm$0.13	&\\
29  &00:14:36.56	&-38:59:10.4   &1.12  &\bf{11.4$\pm$1.4}  &\bf{4.6$\pm$0.9}   &1.0$\pm$0.6        &&\bf{4.7$\pm$0.6}  &\bf{3.5$\pm$0.5}   &0.3$\pm$0.3        &-0.28$\pm$0.09 &-0.78$\pm$0.17 &2.97$\pm$0.30	&\\
30$^*$  &00:14:37.18	&-39:11:22.0   &2.26   &--         &--         &--         &&\bf{0.9$\pm$0.2}  &0.0$\pm$0.1        &0.1$\pm$0.1        &-0.92$\pm$0.14 &0.53 $\pm$0.76 &0.52$\pm$0.17      			&1.98\\
31  &00:14:37.19	&-39:16:15.4   &1.46   &\bf{1.6$\pm$0.4}   &\bf{1.6$\pm$0.3}   &0.5$\pm$0.3        &&--            &--         &--         &-0.03$\pm$0.16 &-0.49$\pm$0.22 &0.70$\pm$0.17      			&\\
32  &00:14:37.31	&-39:06:52.4   &1.38   &0.0$\pm$0.1        &0.2$\pm$0.2        &\bf{1.3$\pm$0.3}   &&0.0$\pm$0.1       &0.3$\pm$0.1        &\bf{1.0$\pm$0.2}   &1.00$\pm$0.57  &0.65 $\pm$0.22 &0.96$\pm$0.16     &\\
33  &00:14:38.39	&-39:15:19.7   &0.83   &\bf{2.7$\pm$0.4}   &\bf{1.6$\pm$0.3}   &\bf{1.1$\pm$0.3}   &&\bf{1.1$\pm$0.2}  &\bf{0.7$\pm$0.1}   &\bf{0.6$\pm$0.1}   &-0.22$\pm$0.11 &-0.14$\pm$0.14 &1.31$\pm$0.13     &\\
34$^*$  &00:14:38.54	&-39:12:41.3   &1.03   &0.7$\pm$0.3        &\bf{1.8$\pm$0.4}   &0.7$\pm$0.3        &&0.5$\pm$0.1       &\bf{0.8$\pm$0.2}   &\bf{0.7$\pm$0.2}   &0.33$\pm$0.17  &-0.25$\pm$0.17 &0.95$\pm$0.14     &3.62\\
35$^*$  &00:14:40.88	&-39:09:38.6   &1.42   &--         &--         &--         &&0.5$\pm$0.1       &\bf{0.9$\pm$0.2}   &\bf{0.7$\pm$0.2}   &0.33$\pm$0.16  &-0.14$\pm$0.16 &1.58$\pm$0.27      			&5.98\\
36  &00:14:42.95	&-39:06:31.0   &0.85   &0.0$\pm$0.2        &\bf{2.9$\pm$0.5}   &\bf{3.3$\pm$0.5}   &&0.2$\pm$0.1       &\bf{1.6$\pm$0.3}   &\bf{2.3$\pm$0.3}   &0.90$\pm$0.13  &0.13 $\pm$0.11 &2.84$\pm$0.25     &\\
37$^*$  &00:14:43.74	&-39:12:41.5   &1.49   &0.2$\pm$0.3        &\bf{1.2$\pm$0.3}   &0.9$\pm$0.3        &&0.1$\pm$0.1       &\bf{0.7$\pm$0.2}   &0.3$\pm$0.2        &0.72$\pm$0.31  &-0.25$\pm$0.22 &0.73$\pm$0.14     &2.78\\
38$^*$  &00:14:44.62	&-39:11:35.9   &0.58   &\bf{6.7$\pm$0.6}   &\bf{10.1$\pm$0.7}  &\bf{8.7$\pm$0.6}   &&\bf{2.1$\pm$0.2}  &\bf{4.8$\pm$0.4}   &\bf{4.3$\pm$0.3}   &0.30$\pm$0.05  &-0.07$\pm$0.05 &7.68$\pm$0.30     &29.1\\
39  &00:14:45.73	&-39:14:35.3   &0.51   &\bf{198.5$\pm$2.9} &\bf{64.0$\pm$1.6}  &\bf{9.8$\pm$0.7}   &&\bf{51.2$\pm$1.0} &\bf{25.1$\pm$0.7}  &\bf{3.6$\pm$0.3}   &-0.45$\pm$0.01 &-0.73$\pm$0.02 &35.9$\pm$0.4   	&\\
40  &00:14:49.00	&-39:22:29.9   &0.82   &1.8$\pm$0.6        &\bf{6.3$\pm$1.0}   &\bf{2.3$\pm$0.7}   &&\bf{1.0$\pm$0.2}  &\bf{2.3$\pm$0.2}   &\bf{2.1$\pm$0.2}   &0.45$\pm$0.08  &-0.11$\pm$0.10 &3.42$\pm$0.27     &\\
41  &00:14:50.55	&-39:09:01.0   &0.62  &\bf{8.8$\pm$0.7}   &\bf{4.9$\pm$0.5}   &\bf{2.7$\pm$0.4}   &&\bf{3.1$\pm$0.5}  &\bf{3.0$\pm$0.4}   &\bf{2.5$\pm$0.4}   &-0.23$\pm$0.10 &-0.22$\pm$0.08 &3.47$\pm$0.22    	&\\
42  &00:14:51.52	&-39:24:16.1   &0.88  &--         &--         &--         &&\bf{3.6$\pm$0.4}  &\bf{2.5$\pm$0.3}   &\bf{1.0$\pm$0.2}   &-0.19$\pm$0.08 &-0.42$\pm$0.11 &4.06$\pm$0.39      			&\\
43$^*$  &00:14:52.02	&-39:10:45.2   &0.50  &\bf{106.0$\pm$2.1} &\bf{119.6$\pm$2.3} &\bf{53.3$\pm$1.6}  &&\bf{32.3$\pm$0.8} &\bf{56.5$\pm$1.1}  &\bf{25.6$\pm$0.8}  &0.16$\pm$0.01  &-0.38$\pm$0.01 &64.0$\pm$0.7   	&242.8\\
44$^*$  &00:14:52.68	&-39:12:23.9   &1.26  &--         &--         &--         &&\bf{1.0$\pm$0.2}  &\bf{0.8$\pm$0.2}   &0.4$\pm$0.2        &-0.11$\pm$0.17 &-0.30$\pm$0.20 &1.39$\pm$0.26      			&5.26\\
45  &00:14:54.54	&-39:04:07.4   &0.85  &\bf{5.6$\pm$0.8}   &\bf{3.2$\pm$0.6}   &0.5$\pm$0.3        &&\bf{2.6$\pm$0.4}  &\bf{2.3$\pm$0.3}   &0.2$\pm$0.2        &-0.16$\pm$0.10 &-0.77$\pm$0.16 &1.70$\pm$0.18    	&\\
46  &00:14:56.04	&-39:16:17.9   &0.84   &\bf{3.0$\pm$0.4}   &\bf{1.9$\pm$0.3}   &1.1$\pm$0.3        &&0.5$\pm$0.1       &\bf{0.8$\pm$0.1}   &\bf{0.4$\pm$0.1}   &-0.04$\pm$0.11 &-0.27$\pm$0.13 &1.15$\pm$0.11	&\\
47$^*$  &00:14:57.00	&-39:11:39.2   &0.51   &\bf{20.3$\pm$1.0}  &\bf{42.2$\pm$1.3}  &\bf{37.2$\pm$1.3}  &&\bf{6.3$\pm$0.4}  &\bf{20.0$\pm$0.6}  &\bf{17.8$\pm$0.6}  &0.44$\pm$0.02  &-0.05$\pm$0.02 &31.6$\pm$0.6   	&119.7\\
48$^*$  &00:14:57.04	&-39:12:26.8   &1.54   &\bf{2.6$\pm$0.6}   &1.3$\pm$0.4        &0.2$\pm$0.2        &&0.8$\pm$0.2       &0.5$\pm$0.2        &0.1$\pm$0.1        &-0.29$\pm$0.20 &-0.75$\pm$0.32 &0.59$\pm$0.12     &2.25\\
49  &00:14:58.48	&-39:22:11.1   &0.69   &\bf{7.0$\pm$0.8}   &\bf{2.2$\pm$0.4}   &\bf{1.7$\pm$0.3}   &&\bf{1.1$\pm$0.2}  &\bf{1.4$\pm$0.2}   &\bf{0.6$\pm$0.1}   &-0.30$\pm$0.08 &-0.26$\pm$0.12 &1.99$\pm$0.16     &\\
50  &00:14:59.39	&-39:17:24.9   &1.32   &0.3$\pm$0.3        &0.8$\pm$0.3        &\bf{1.4$\pm$0.4}   &&0.0$\pm$0.0       &\bf{0.5$\pm$0.1}   &\bf{0.5$\pm$0.1}   &0.87$\pm$0.17  &0.12 $\pm$0.20 &0.91$\pm$0.14    	&\\
51$^*$  &00:15:00.62	&-39:12:16.4   &1.16   &--         &--         &--         &&0.4$\pm$0.1       &\bf{0.7$\pm$0.1}   &\bf{0.8$\pm$0.1}   &0.31$\pm$0.15  &-0.04$\pm$0.12 &1.74$\pm$0.22      			&6.58\\
52$^*$  &00:15:01.18	&-39:12:40.6   &0.98   &\bf{5.0$\pm$0.6}   &\bf{1.6$\pm$0.3}   &0.2$\pm$0.2        &&\bf{1.4$\pm$0.2}  &0.5$\pm$0.2        &0.2$\pm$0.1        &-0.49$\pm$0.12 &-0.72$\pm$0.22 &0.92$\pm$0.11     &3.49\\
53$^*$  &00:15:03.79	&-39:14:59.8   &1.52   &\bf{2.1$\pm$0.3}   &0.2$\pm$0.1        &0.1$\pm$0.2        &&--            &--         &--         &-0.84$\pm$0.11 &-0.18$\pm$0.73 &0.30$\pm$0.10      			&1.12\\
54$^*$  &00:15:04.01	&-39:11:53.3   &0.67   &0.9$\pm$0.3        &\bf{4.0$\pm$0.4}   &\bf{3.6$\pm$0.4}   &&0.2$\pm$0.1       &\bf{1.6$\pm$0.2}   &\bf{1.5$\pm$0.2}   &0.68$\pm$0.11  &-0.04$\pm$0.08 &2.80$\pm$0.18     &10.6\\
55$^*$  &00:15:04.47	&-39:10:45.9   &0.83   &\bf{2.2$\pm$0.4}   &\bf{1.6$\pm$0.3}   &0.9$\pm$0.3        &&\bf{0.5$\pm$0.1}  &\bf{1.0$\pm$0.2}   &\bf{0.7$\pm$0.1}   &0.09$\pm$0.12  &-0.19$\pm$0.11 &1.22$\pm$0.12     &4.63\\
56  &00:15:07.98	&-39:16:55.8   &0.91   &0.4$\pm$0.2        &\bf{2.4$\pm$0.3}   &\bf{3.1$\pm$0.4}   &&0.0$\pm$0.0       &\bf{0.9$\pm$0.1}   &\bf{1.5$\pm$0.2}   &0.87$\pm$0.12  &0.19$\pm$0.09  &2.28$\pm$0.17     &\\
57  &00:15:08.77	&-39:02:29.1   &1.39   &\bf{1.9$\pm$0.6}   &\bf{1.3$\pm$0.4}   &0.2$\pm$0.3        &&\bf{2.0$\pm$0.4}  &\bf{1.2$\pm$0.3}   &0.4$\pm$0.3        &-0.24$\pm$0.17 &-0.59$\pm$0.30 &0.69$\pm$0.16 	&\\
58$^*$  &00:15:13.44	&-39:12:45.0   &2.49  &--         &--         &--         &&\bf{0.7$\pm$0.2}  &0.1 $\pm$0.1       &0.0$\pm$0.1        &-0.69$\pm$0.20 &-0.50$\pm$0.64 &0.39$\pm$0.12      			&1.47\\
59  &00:15:16.42	&-39:20:59.0   &0.87   &\bf{4.9$\pm$0.6}   &\bf{2.7$\pm$0.4}   &\bf{2.0$\pm$0.4}   &&\bf{1.2$\pm$0.2}  &\bf{0.9$\pm$0.2}   &\bf{0.7$\pm$0.2}   &-0.25$\pm$0.13 &-0.13$\pm$0.14 &1.97$\pm$0.19     &\\
60  &00:15:19.57	&-39:09:17.6   &1.48   &\bf{2.3$\pm$0.5}   &0.9$\pm$0.3        &0.3$\pm$0.3        &&--            &--         &--         &-0.42$\pm$0.14 &-0.52$\pm$0.33 &0.54$\pm$0.15      			&\\
61  &00:15:19.97	&-39:19:07.2   &1.45   &--         &--         &--         &&\bf{0.5$\pm$0.1}  &\bf{0.5$\pm$0.1}   &0.3$\pm$0.1        &-0.06$\pm$0.16 &-0.25$\pm$0.21 &0.82$\pm$0.17      			&\\
62  &00:15:21.94	&-39:18:43.5   &0.99  &--         &--         &--         &&\bf{0.4$\pm$0.1}  &\bf{1.5$\pm$0.2}   &\bf{1.1$\pm$0.2}   &0.61$\pm$0.10  &-0.18$\pm$0.10 &2.33$\pm$0.25      			&\\
63$^*$  &00:15:22.83	&-39:12:41.1   &1.31   &\bf{3.4$\pm$0.8}   &0.0$\pm$0.2        &0.0$\pm$0.1        &&--            &--         &--             &-1.00$\pm$0.09 &--     &0.35$\pm$0.10      			&1.32\\
64$^*$  &00:15:22.93	&-39:13:30.0   &2.70  &--         &--         &--         &&0.3$\pm$0.1       &\bf{0.8$\pm$0.2}   &0.2$\pm$0.1        &0.46$\pm$0.16  &-0.64$\pm$0.19 &0.72$\pm$0.17      			&2.72\\
65  &00:15:23.64	&-39:09:47.2   &1.13   &0.6$\pm$0.3        &\bf{1.5$\pm$0.3}   &\bf{1.3$\pm$0.3}   &&0.2$\pm$0.1       &\bf{0.7$\pm$0.2}   &\bf{0.8$\pm$0.2}   &0.49$\pm$0.20  &-0.01$\pm$0.16 &1.20$\pm$0.15    	&\\
66  &00:15:24.45	&-39:22:04.5   &2.42   &\bf{1.3$\pm$0.3}   &0.0$\pm$0.1        &0.3$\pm$0.2        &&--            &--         &--             &-1.00$\pm$0.21 &1.00 $\pm$0.89 &0.30$\pm$0.14      		&\\
67  &00:15:25.61	&-39:26:02.5   &1.14  &\bf{5.6$\pm$0.8}   &1.5$\pm$0.5        &0.2$\pm$0.3        &&\bf{1.1$\pm$0.2}  &\bf{0.8$\pm$0.2}   &0.3$\pm$0.2        &-0.43$\pm$0.15 &-0.51$\pm$0.30 &1.05$\pm$0.16     	&\\
68  &00:15:25.82	&-39:19:40.2   &1.27  &\bf{1.8$\pm$0.4}   &\bf{0.9$\pm$0.2}   &0.5$\pm$0.3        &&0.2$\pm$0.1       &0.1$\pm$0.1        &0.2$\pm$0.1        &-0.31$\pm$0.24 &-0.10$\pm$0.31 &0.54$\pm$0.11	&\\
69$^*$  &00:15:26.01	&-39:14:05.9   &1.00  &\bf{3.9$\pm$0.5}   &\bf{4.9$\pm$0.4}   &\bf{1.7$\pm$0.4}   &&--            &--         &--         &0.12$\pm$0.08  &-0.48$\pm$0.09 &2.11$\pm$0.22      			&8.00\\
\hline
\end{tabular}
}
\end{center}
\end{table*}
\begin{table*}

\begin{center}
{\scriptsize
\begin{tabular}{lcccccccccccccc}

\hline

SRC     &RA		&DEC    	&r$_1\sigma$    &\multicolumn{3}{c}{pn cts $\rm ks^{-1}$}       &&\multicolumn{3}{c}{MOS cts $\rm ks^{-1}$}                 &HR1       	     &HR2           &{\it f}$_X$            &L$_X$  \\ \cline{5-7} \cline{9-11}
ID  	&(hh:mm:ss)  &(\deg:\arcmin:\arcsec)&(\arcsec) &S       &M     		    &H          	&&S         	   &M          		&H          	    &       	     &              &($\times 10^{-14}$)    &($\times 10^{36}$)\\
\hline
70  &00:15:27.91	&-39:05:07.7   &1.28  &\bf{2.8$\pm$0.5}   &0.3$\pm$0.2        &0.1$\pm$0.2        &&0.5$\pm$0.2       &0.4$\pm$0.2        &0.0$\pm$0.0        &-0.68$\pm$0.14 &-0.95$\pm$0.61 &0.40$\pm$0.08      &\\
71$^*$  &00:15:28.87	&-39:13:18.7   &0.50  &\bf{833.6$\pm$5.9} &\bf{772.8$\pm$5.8} &\bf{225.1$\pm$3.2} &&\bf{110.6$\pm$1.9}    &\bf{158.8$\pm$2.3} &\bf{43.9$\pm$1.2}  &0.04$\pm$0.01  &-0.52$\pm$0.01 &247.3$\pm$1.5  &937.6\\
72  &00:15:30.29	&-39:07:15.0   &0.72   &\bf{2.3$\pm$0.5}   &\bf{2.7$\pm$0.5}   &\bf{4.4$\pm$0.6}   &&0.6$\pm$0.1       &\bf{1.5$\pm$0.2}   &\bf{3.1$\pm$0.3}   &0.28$\pm$0.13  &0.32 $\pm$0.08 &4.04$\pm$0.27     &\\
73$^*$  &00:15:33.18	&-39:17:35.3   &1.13  &1.0$\pm$0.3        &\bf{1.0$\pm$0.2}   &0.7$\pm$0.2        &&\bf{0.4$\pm$0.1}  &0.2$\pm$0.1        &0.2$\pm$0.1        &-0.12$\pm$0.21 &-0.10$\pm$0.26 &0.62$\pm$0.10      &2.33\\
74$^*$  &00:15:33.98	&-39:15:12.2   &0.67   &\bf{9.9$\pm$0.7}   &\bf{2.7$\pm$0.4}   &0.6$\pm$0.3        &&\bf{2.0$\pm$0.2}  &\bf{1.4$\pm$0.2}   &0.0$\pm$0.1        &-0.44$\pm$0.06 &-0.87$\pm$0.13 &1.59$\pm$0.11     &6.02\\
75$^*$  &00:15:34.27	&-39:14:24.8   &0.52  &\bf{27.6$\pm$1.1}  &\bf{35.1$\pm$1.2}  &\bf{30.0$\pm$1.1}  &&\bf{6.2$\pm$0.3}  &\bf{13.2$\pm$0.5}  &\bf{11.7$\pm$0.5}  &0.24$\pm$0.03  &-0.07$\pm$0.03 &24.5$\pm$0.5   	&93.0\\
76  &00:15:34.80	&-39:21:10.4   &1.00   &\bf{2.1$\pm$0.4}   &\bf{2.5$\pm$0.4}   &\bf{2.7$\pm$0.4}   &&\bf{0.3$\pm$0.1}  &\bf{0.8$\pm$0.1}   &\bf{0.3$\pm$0.1}   &0.29$\pm$0.13  &-0.19$\pm$0.11 &1.13$\pm$0.11     &\\
77  &00:15:37.13	&-39:09:42.7   &1.35   &\bf{2.3$\pm$0.4}   &0.5$\pm$0.2        &0.1$\pm$0.2        &&--            &--         &--         &-0.67$\pm$0.10 &-0.82$\pm$0.39 &0.37$\pm$0.10      			&\\
78  &00:15:37.43	&-39:19:54.1   &1.58   &0.4$\pm$0.3        &\bf{1.0$\pm$0.2}   &0.2$\pm$0.2        &&\bf{0.4$\pm$0.1}  &0.2$\pm$0.1        &0.0$\pm$0.0        &-0.02$\pm$0.28 &-0.77$\pm$0.36 &0.26$\pm$0.06     &\\
79$^*$  &00:15:38.00	&-39:12:32.8   &0.69   &\bf{3.7$\pm$0.4}   &\bf{3.1$\pm$0.4}   &\bf{1.4$\pm$0.3}   &&\bf{1.1$\pm$0.1}  &\bf{1.4$\pm$0.2}   &\bf{0.9$\pm$0.1}   &0.02$\pm$0.08  &-0.27$\pm$0.09 &1.91$\pm$0.13     &7.22\\
80$^*$  &00:15:38.54	&-39:15:22.5   &1.71  &\bf{1.3$\pm$0.3}   &\bf{2.3$\pm$0.4}   &\bf{1.2$\pm$0.3}   &&--            &--         &--         &0.27$\pm$0.14  &-0.31$\pm$0.13 &1.17$\pm$0.18      			&4.42\\
81  &00:15:39.24	&-39:26:44.9   &0.83  &\bf{8.9$\pm$1.0}   &\bf{3.3$\pm$0.6}   &\bf{3.0$\pm$0.6}   &&\bf{2.0$\pm$0.3}  &\bf{1.2$\pm$0.2}   &\bf{1.4$\pm$0.3}   &-0.38$\pm$0.09 &0.01 $\pm$0.14 &3.15$\pm$0.26      &\\
82  &00:15:39.41	&-39:00:59.0   &1.50  &\bf{5.4$\pm$1.0}   &\bf{4.9$\pm$0.8}   &1.5$\pm$0.7        &&--            &--         &--         &-0.05$\pm$0.13 &-0.53$\pm$0.18 &2.13$\pm$0.43      			&\\
83  &00:15:40.45	&-39:26:06.5   &1.25   &\bf{2.4$\pm$0.6}   &\bf{1.5$\pm$0.4}   &0.8$\pm$0.4        &&\bf{0.7$\pm$0.2}  &\bf{0.6$\pm$0.2}   &0.3$\pm$0.1        &-0.15$\pm$0.17 &-0.31$\pm$0.25 &0.98$\pm$0.16     &\\
84$^*$  &00:15:41.26	&-39:15:09.3   &1.15   &\bf{3.8$\pm$0.6}   &0.7$\pm$0.3        &0.0$\pm$0.1        &&\bf{0.7$\pm$0.1}  &0.2$\pm$0.1        &0.0$\pm$0.0        &-0.66$\pm$0.14 &-0.98$\pm$0.33 &0.42$\pm$0.06     &1.58\\
85$^*$  &00:15:43.82	&-39:16:41.8   &2.27  &\bf{1.4$\pm$0.3}   &0.1$\pm$0.1        &0.0$\pm$0.1        &&--            &--         &--             &-0.93$\pm$0.13 &-1.00$\pm$3.57 &0.15$\pm$0.06      		&0.58\\
86  &00:15:44.17	&-39:20:43.5   &0.94  &\bf{2.8$\pm$0.4}   &\bf{1.5$\pm$0.3}   &0.7$\pm$0.2        &&\bf{0.5$\pm$0.1}  &\bf{0.4$\pm$0.1}   &0.2$\pm$0.1        &-0.22$\pm$0.14 &-0.37$\pm$0.17 &0.78$\pm$0.10    	&\\
87  &00:15:44.28	&-39:12:09.2   &0.72   &\bf{6.6$\pm$1.2}   &\bf{3.1$\pm$0.8}   &2.7$\pm$0.8        &&\bf{1.7$\pm$0.2}  &\bf{1.8$\pm$0.2}   &\bf{1.0$\pm$0.1}   &-0.06$\pm$0.14 &-0.27$\pm$0.17 &2.79$\pm$0.19     &\\
88  &00:15:44.79	&-39:00:30.4   &1.09  &\bf{3.7$\pm$0.9}   &\bf{5.5$\pm$1.0}   &2.4$\pm$0.7        &&--            &--         &--         &0.20$\pm$0.15  &-0.39$\pm$0.15 &2.58$\pm$0.44      			&\\
89  &00:15:45.87	&-39:10:23.0   &0.96  &\bf{2.5$\pm$0.4}   &\bf{1.0$\pm$0.2}   &\bf{0.9$\pm$0.3}   &&0.3$\pm$0.1       &\bf{0.4$\pm$0.1}   &0.3$\pm$0.1        &-0.20$\pm$0.12 &-0.09$\pm$0.17 &0.84$\pm$0.11      &\\
90  &00:15:48.35	&-39:18:39.3   &0.59  &\bf{11.9$\pm$0.8}  &\bf{6.3$\pm$0.5}   &\bf{2.8$\pm$0.4}   &&\bf{3.3$\pm$0.3}  &\bf{2.3$\pm$0.2}   &\bf{1.3$\pm$0.2}   &-0.26$\pm$0.06 &-0.33$\pm$0.07 &3.93$\pm$0.19     	&\\
91  &00:15:48.54	&-39:03:42.0   &1.08   &\bf{4.4$\pm$0.7}   &\bf{2.9$\pm$0.6}   &1.0$\pm$0.5        &&\bf{1.7$\pm$0.3}  &\bf{1.1$\pm$0.3}   &0.5$\pm$0.2        &-0.20$\pm$0.14 &-0.43$\pm$0.21 &1.65$\pm$0.23    	&\\
92  &00:15:49.27	&-39:23:48.2   &1.47   &0.9$\pm$0.3        &0.5$\pm$0.3        &0.8$\pm$0.3        &&0.3$\pm$0.1       &\bf{0.5$\pm$0.1}   &0.3$\pm$0.1        &0.11$\pm$0.29  &-0.15$\pm$0.24 &0.67$\pm$0.13 	&\\
93$^*$  &00:15:50.29	&-39:16:36.8   &2.24   &\bf{1.2$\pm$0.3}   &0.0$\pm$0.1        &0.0$\pm$0.1        &&--            &--         &--         &-1.00$\pm$0.18 &--             &0.12$\pm$0.05      			&0.47\\
94  &00:15:51.69	&-39:02:32.5   &1.53   &1.3$\pm$0.5        &1.6$\pm$0.5        &\bf{2.6$\pm$0.6}   &&0.3$\pm$0.2       &\bf{1.1$\pm$0.3}   &0.7$\pm$0.3        &0.39$\pm$0.23  &0.05 $\pm$0.20 &1.69$\pm$0.26     &\\
95  &00:15:51.79	&-39:03:55.4   &1.34   &0.6$\pm$0.5        &\bf{1.7$\pm$0.5}   &\bf{1.9$\pm$0.6}   &&0.1$\pm$0.1       &0.1$\pm$0.1        &\bf{1.2$\pm$0.3}   &0.40$\pm$0.32  &0.46 $\pm$0.20 &1.52$\pm$0.26	&\\
96  &00:15:52.19	&-39:12:32.1   &1.31   &0.9$\pm$0.3        &\bf{1.0$\pm$0.2}   &\bf{0.9$\pm$0.3}   &&--            &--         &--         &0.08$\pm$0.18  &-0.04$\pm$0.18 &0.76$\pm$0.15      			&\\
97  &00:15:52.80	&-39:10:59.8   &0.87   &\bf{3.1$\pm$0.5}   &\bf{1.0$\pm$0.3}   &0.9$\pm$0.3        &&\bf{0.6$\pm$0.1}  &\bf{0.6$\pm$0.1}   &\bf{0.4$\pm$0.1}   &-0.33$\pm$0.11 &-0.17$\pm$0.19 &1.00$\pm$0.11	&\\
98$^*$  &00:15:54.92	&-39:16:06.9   &1.34   &0.5$\pm$0.3        &0.7$\pm$0.2        &0.4$\pm$0.2        &&0.0$\pm$0.1       &0.0$\pm$0.0        &\bf{0.5$\pm$0.1}   &0.16$\pm$0.31  &-0.23$\pm$0.30 &0.52$\pm$0.10     &1.96\\
99 &00:15:55.83		&-39:09:37.8   &1.02   &\bf{3.2$\pm$0.5}   &\bf{1.4$\pm$0.3}   &0.4$\pm$0.2        &&\bf{1.4$\pm$0.4}  &\bf{0.6$\pm$0.2}   &\bf{0.6$\pm$0.2}   &-0.38$\pm$0.19 &-0.38$\pm$0.24 &0.87$\pm$0.13    	&\\
100 &00:15:56.82	&-39:13:39.5   &1.06   &0.1$\pm$0.1        &\bf{0.8$\pm$0.2}   &\bf{2.1$\pm$0.4}   &&0.0$\pm$0.0       &\bf{0.3$\pm$0.1}   &\bf{0.4$\pm$0.1}   &0.97$\pm$0.12  &0.32 $\pm$0.16 &0.84$\pm$0.11    	&\\
101$^*$ &00:15:57.33	&-39:16:12.9   &1.14  &\bf{3.7$\pm$0.5}   &0.3$\pm$0.2        &0.0$\pm$0.1        &&\bf{0.7$\pm$0.1}  &0.0$\pm$0.1        &0.0$\pm$0.0        &-0.87$\pm$0.11 &-1.00$\pm$1.00 &0.38$\pm$0.05      &1.44\\
102 &00:15:57.40	&-39:24:27.9   &0.79  &\bf{5.6$\pm$0.7}   &\bf{2.1$\pm$0.4}   &1.2$\pm$0.4        &&\bf{1.8$\pm$0.2}  &\bf{0.7$\pm$0.2}   &\bf{0.7$\pm$0.2}   &-0.46$\pm$0.10 &-0.13$\pm$0.17 &1.73$\pm$0.16      &\\
103 &00:15:59.24	&-39:03:23.8   &1.00   &\bf{6.8$\pm$1.0}   &\bf{2.2$\pm$0.6}   &1.3$\pm$0.5        &&\bf{2.2$\pm$0.4}  &0.8$\pm$0.2        &0.7$\pm$0.2        &-0.50$\pm$0.14 &-0.13$\pm$0.22 &1.94$\pm$0.24     &\\
104 &00:15:59.52	&-39:10:18.1   &1.56   &\bf{1.9$\pm$0.4}   &\bf{1.1$\pm$0.3}   &0.7$\pm$0.3        &&--            &--         &--         &-0.27$\pm$0.15 &-0.23$\pm$0.22 &0.76$\pm$0.16      			&\\
105 &00:15:59.54	&-39:19:07.2   &0.74  &\bf{4.1$\pm$0.5}   &\bf{2.8$\pm$0.4}   &\bf{1.3$\pm$0.3}   &&\bf{0.9$\pm$0.2}  &\bf{1.0$\pm$0.1}   &\bf{0.6$\pm$0.1}   &-0.09$\pm$0.10 &-0.34$\pm$0.12 &1.55$\pm$0.13      &\\
106$^*$ &00:16:00.50	&-39:16:55.0   &1.45  &0.0$\pm$0.1        &0.5$\pm$0.2        &\bf{1.7$\pm$0.3}   &&0.0$\pm$0.0       &0.1$\pm$0.1        &\bf{0.4$\pm$0.1}   &0.85$\pm$0.51  &0.55 $\pm$0.21 &0.73$\pm$0.11      &2.78\\
107 &00:16:01.26	&-39:21:38.3   &1.61  &0.4$\pm$0.3        &0.5$\pm$0.2        &0.9$\pm$0.3        &&0.0$\pm$0.1       &0.2$\pm$0.1        &\bf{0.4$\pm$0.1}   &0.50$\pm$0.38  &0.26 $\pm$0.25 &0.62$\pm$0.11      &\\
108$^*$ &00:16:01.75	&-39:16:23.1   &0.75  &0.2$\pm$0.1        &0.5$\pm$0.2        &\bf{5.1$\pm$0.5}   &&0.0$\pm$0.0       &0.1$\pm$0.1        &\bf{2.1$\pm$0.2}   &0.57$\pm$0.40  &0.86 $\pm$0.06 &2.93$\pm$0.21      &11.1\\
109 &00:16:02.79	&-39:12:11.5   &0.62  &\bf{9.8$\pm$0.7}   &\bf{4.9$\pm$0.5}   &\bf{2.4$\pm$0.4}   &&\bf{2.0$\pm$0.2}  &\bf{2.0$\pm$0.2}   &\bf{1.3$\pm$0.2}   &-0.21$\pm$0.07 &-0.27$\pm$0.08 &3.27$\pm$0.18      &\\
110 &00:16:04.47	&-39:11:39.4   &0.63   &\bf{6.3$\pm$0.6}   &\bf{5.2$\pm$0.5}   &\bf{2.7$\pm$0.4}   &&\bf{2.3$\pm$0.3}  &\bf{2.1$\pm$0.3}   &\bf{1.6$\pm$0.2}   &-0.07$\pm$0.08 &-0.24$\pm$0.09 &3.34$\pm$0.20     &\\
111 &00:16:05.24	&-39:05:31.8   &1.05  &1.3$\pm$0.4        &\bf{3.2$\pm$0.6}   &\bf{2.2$\pm$0.5}   &&0.4$\pm$0.2       &\bf{1.8$\pm$0.3}   &\bf{1.3$\pm$0.3}   &0.52$\pm$0.15  &-0.18$\pm$0.14 &2.15$\pm$0.25      &\\
112$^*$ &00:16:06.67	&-39:15:22.5   &0.86  &\bf{1.2$\pm$0.3}   &\bf{2.0$\pm$0.3}   &\bf{2.6$\pm$0.4}   &&0.2$\pm$0.1       &\bf{0.9$\pm$0.2}   &\bf{1.0$\pm$0.2}   &0.41$\pm$0.16  &0.08 $\pm$0.12 &1.85$\pm$0.17      &7.03\\
113$^*$ &00:16:08.34	&-39:17:41.1   &0.87  &\bf{2.1$\pm$0.4}   &\bf{3.4$\pm$0.5}   &\bf{2.1$\pm$0.4}   &&\bf{0.5$\pm$0.1}  &\bf{1.0$\pm$0.2}   &\bf{0.3$\pm$0.1}   &0.27$\pm$0.12  &-0.35$\pm$0.12 &1.32$\pm$0.13      &5.02\\
114 &00:16:08.81	&-39:11:22.2   &1.15  &\bf{1.7$\pm$0.4}   &\bf{0.8$\pm$0.2}   &0.7$\pm$0.3        &&\bf{0.5$\pm$0.1}  &\bf{0.5$\pm$0.1}   &0.2$\pm$0.1        &-0.23$\pm$0.16 &-0.19$\pm$0.23 &0.70$\pm$0.11      &\\
115 &00:16:09.59	&-39:07:45.4   &1.90   &\bf{1.9$\pm$0.4}   &\bf{1.2$\pm$0.3}   &0.0$\pm$0.2        &&--            &--         &--         &-0.24$\pm$0.16 &-1.00$\pm$0.28 &0.37$\pm$0.11      			&\\
116 &00:16:10.33	&-39:01:25.6   &2.45   &\bf{2.4$\pm$0.6}   &0.8$\pm$0.5        &1.5$\pm$0.6        &&--            &--         &--             &-0.48$\pm$0.24 &0.28 $\pm$0.32 &1.20$\pm$0.36      		&\\
117$^*$ &00:16:10.37	&-39:17:05.3   &2.80  &1.0$\pm$0.4        &\bf{1.2$\pm$0.3}   &0.0$\pm$0.1        &&--            &--         &--         &0.06$\pm$0.22  &-0.92$\pm$0.19 &0.31$\pm$0.09      			&1.18\\
118 &00:16:10.81	&-39:20:13.7   &0.76   &\bf{4.5$\pm$0.6}   &\bf{3.4$\pm$0.5}   &\bf{1.9$\pm$0.4}   &&\bf{0.6$\pm$0.1}  &\bf{1.0$\pm$0.2}   &\bf{1.0$\pm$0.2}   &-0.02$\pm$0.13 &-0.16$\pm$0.13 &2.06$\pm$0.18     &\\
119 &00:16:12.59	&-39:03:45.0   &2.79   &\bf{2.1$\pm$0.6}   &0.3$\pm$0.4        &0.8$\pm$0.5        &&--            &--             &--         &-0.74$\pm$0.26 &0.45 $\pm$0.50 &0.73$\pm$0.28      		&\\
120 &00:16:14.28	&-39:27:27.3   &1.62   &--         &--         &--         &&0.5$\pm$0.2       &\bf{0.9$\pm$0.2}   &\bf{0.8$\pm$0.2}   &0.24$\pm$0.23  &-0.02$\pm$0.18 &1.73$\pm$0.33      			&\\
121 &00:16:14.39	&-39:08:48.1   &1.24   &0.6$\pm$0.3        &\bf{1.5$\pm$0.3}   &0.6$\pm$0.3        &&\bf{0.6$\pm$0.1}  &0.4$\pm$0.1        &0.3$\pm$0.1        &0.15$\pm$0.22  &-0.32$\pm$0.24 &0.72$\pm$0.14     &\\
122$^*$ &00:16:19.34	&-39:16:45.5   &2.61   &0.2$\pm$0.2        &\bf{1.3$\pm$0.3}   &0.5$\pm$0.3        &&--            &--         &--         &0.75$\pm$0.26  &-0.48$\pm$0.24 &0.46$\pm$0.15      			&1.76\\
123 &00:16:23.06	&-39:23:53.7   &1.75   &\bf{2.2$\pm$0.5}   &0.9$\pm$0.3        &0.4$\pm$0.3        &&--            &--         &--         &-0.43$\pm$0.18 &-0.42$\pm$0.37 &0.57$\pm$0.18      			&\\
124 &00:16:27.39	&-39:09:26.6   &1.66   &1.4$\pm$0.4        &\bf{1.6$\pm$0.4}   &0.7$\pm$0.4        &&--            &--         &- -            &0.08$\pm$0.20  &-0.42$\pm$0.25 &0.75$\pm$0.21      		&\\
125 &00:16:33.07	&-39:05:32.6   &0.73   &\bf{13.6$\pm$1.4}  &\bf{6.1$\pm$1.0}   &\bf{3.7$\pm$0.7}   &&\bf{4.9$\pm$0.6}  &\bf{3.3$\pm$0.5}   &\bf{1.4$\pm$0.4}   &-0.30$\pm$0.08 &-0.33$\pm$0.12 &4.72$\pm$0.36     &\\
126 &00:16:35.81	&-39:17:24.4   &1.39   &\bf{2.0$\pm$0.4}   &0.0$\pm$0.1        &0.3$\pm$0.3        &&--            &--         &--         &-1.00$\pm$0.08 &1.00 $\pm$0.53 &0.36$\pm$0.15      			&\\
127 &00:16:36.45	&-39:15:01.5   &2.17   &--         &--             &--         &&0.1$\pm$0.1       &\bf{0.7$\pm$0.2}   &0.0$\pm$0.1        &0.78$\pm$0.22  &-1.00$\pm$0.22 &0.35$\pm$0.14      			&\\
128 &00:16:36.48	&-39:25:14.6   &1.45   &--         &--             &--         &&\bf{1.9$\pm$0.4}  &\bf{0.9$\pm$0.3}   &0.8$\pm$0.2        &-0.36$\pm$0.15 &-0.04$\pm$0.20 &2.33$\pm$0.38      			&\\
129 &00:16:37.43	&-39:06:33.3   &1.47   &\bf{3.8$\pm$0.9}   &\bf{2.1$\pm$0.6}   &1.2$\pm$0.6        &&--            &--         &--         &-0.29$\pm$0.17 &-0.29$\pm$0.26 &1.37$\pm$0.35      			&\\
130 &00:16:38.59	&-39:16:24.1   &1.42   &\bf{2.3$\pm$0.5}   &\bf{1.9$\pm$0.4}   &0.6$\pm$0.3        &&--            &--         &--         &-0.11$\pm$0.15 &-0.54$\pm$0.22 &0.84$\pm$0.20      			&\\
131 &00:16:40.12	&-39:11:35.3   &1.19   &\bf{3.4$\pm$0.6}   &\bf{2.2$\pm$0.5}   &0.8$\pm$0.4        &&\bf{1.0$\pm$0.2}  &\bf{0.9$\pm$0.2}   &0.3$\pm$0.2        &-0.15$\pm$0.15 &-0.47$\pm$0.23 &1.16$\pm$0.19     &\\
132 &00:16:42.18	&-39:18:20.2   &0.70   &\bf{11.0$\pm$1.1}  &\bf{7.3$\pm$0.9}   &\bf{5.0$\pm$0.8}   &&\bf{2.3$\pm$0.3}  &\bf{3.4$\pm$0.4}   &\bf{2.9$\pm$0.4}   &-0.04$\pm$0.08 &-0.14$\pm$0.09 &5.54$\pm$0.35     &\\
133 &00:16:43.46	&-39:08:26.8   &1.84   &0.2$\pm$0.3        &\bf{2.0$\pm$0.5}   &1.3$\pm$0.5        &&--            &--         &--         &0.86$\pm$0.28  &-0.23$\pm$0.22 &1.02$\pm$0.29      			&\\
134 &00:16:47.41	&-39:15:02.9   &1.13   &\bf{4.0$\pm$0.7}   &\bf{3.2$\pm$0.6}   &0.9$\pm$0.5        &&\bf{1.2$\pm$0.3}  &\bf{1.0$\pm$0.3}   &\bf{1.6$\pm$0.3}   &-0.10$\pm$0.16 &-0.11$\pm$0.17 &1.82$\pm$0.24     &\\
135 &00:16:48.71	&-39:09:32.0   &0.97   &\bf{6.0$\pm$1.0}   &\bf{4.3$\pm$0.8}   &1.1$\pm$0.5        &&\bf{1.5$\pm$0.4}  &\bf{1.3$\pm$0.3}   &\bf{1.1$\pm$0.3}   &-0.14$\pm$0.18 &-0.37$\pm$0.17 &2.14$\pm$0.27     &\\
136 &00:16:55.54	&-39:17:50.6   &1.11   &\bf{6.1$\pm$0.9}   &\bf{3.9$\pm$0.7}   &\bf{2.4$\pm$0.7}   &&\bf{2.1$\pm$0.4}  &\bf{1.3$\pm$0.3}   &\bf{1.7$\pm$0.4}   &-0.23$\pm$0.14 &-0.06$\pm$0.16 &2.93$\pm$0.32     &\\
137 &00:17:04.21	&-39:10:43.4   &1.09   &\bf{12.6$\pm$1.6}  &\bf{5.9$\pm$1.0}   &2.7$\pm$0.8        &&--            &--             &--         &-0.36$\pm$0.09 &-0.36$\pm$0.15 &3.69$\pm$0.52      		&\\

\hline
\end{tabular}
\begin{minipage}[t]{7in}
{\sc Notes:} (1) Source number (objects within the \d25 ellipse are
marked by *); (2) RA in hh:mm:ss (J2000 coordinates); (3) DEC in
\deg:\arcmin:\arcsec (J2000 coordinates); (4) 1$\sigma$ error radius
including a 0.5\arcsec systematic error; (5,6 \&7) and (8,9 \& 10)
Source count rates in the soft (0.3--1\,keV), medium (1--2\,keV) \&
hard (2--6\,keV) bands for the pn and MOS cameras respectively, with
significant source detections ($>4\sigma$) highlighted in bold; (11 \&
12) Soft (HR1) and hard (HR2) hardness ratios (as defined in the
text); (13) Source X-ray flux (0.3--6\,keV) in units of $10^{-14}
\ergcms$; (14) Source X-ray luminosity (0.3--6\,keV) in units of
$10^{38} \ergsec$ (assuming a distance to NGC 55 of 1.78\,Mpc).
\end{minipage}
}
\end{center}
\end{table*}


The full X-ray source catalogue appears in Table~\ref{allsrcs}.  The table
provides the following information: \\
{\it column 1:} The source identification number - sources located within the
optical \d25 ellipse are marked with an asterisk. \\
{\it column 2-3:} The source RA and DEC positions (J2000).  The
correct IAU designated title for each source can be produced by
truncating the coordinates in these columns.  For example, Source 1
becomes XMMU J001328.3-390903.\\
{\it column 4:} The 1$\sigma$ uncertainty in the source position plus a 0.5\arcsec systematic error.\\
{\it column 5-7:} The measured count rate (corrected for the
vignetting) in the pn camera in the soft (S), medium (M) and hard (H)
bands. For sources detected in both observations, these count rates
are the weighted mean values from the two exposures.\\
{\it column 8-10:} The measured count rate (corrected for the
vignetting) in the MOS cameras in the S, M \& H bands. Where
appropriate, weighted mean values are quoted as for the pn data.
Values are quoted for one MOS camera.\\
{\it column 11-12:} Two hardness ratios calculated as
$HR1=(M-S)/(M+S)$ and $HR2=(H-M)/(H+M)$.  For sources
detected in more than one instrument or exposure we quote the weighted
mean hardness ratios.\\
{\it column 13:} The measured flux in the 0.3--6 keV energy range. The count 
rate in each detection band was converted to an X-ray
flux using Energy Conversion Factors (ECFs) calculated for each
instrument from a power-law continuum spectrum with $\Gamma = 1.7$
absorbed by the Galactic foreground column towards NGC 55 (\nh$ =
1.55 \times 10^{20}$ cm$^{-2}$; \citealt{stark92}). Again, for
sources detected in more than one instrument or exposure we took
the weighted mean flux values for the particular band. The fluxes
derived from the individual bands were subsequently added to
provide the quoted instrumental flux (0.3--6 keV).\\ 
{\it column 14:} The derived X-ray luminosities for
sources within the \d25 ellipse, assuming a distance of 1.78 Mpc. \\


\subsection{X-ray sources associated with NGC 55}

Table \ref{sigdet} gives a breakdown of the number of detected sources
as a function of instrument and energy band.  The \d25 ellipse of the
galaxy occupies $\sim$13 per cent of the total field of view, but
encompasses $\sim$30 per cent of the sources.  We have carried out a
Monte Carlo simulation to investigate whether the 42 sources detected
within the \d25 ellipse of NGC 55, represent an excess compared to the
number of background objects expected by chance. Our predictions are
based on the log {\it N} -- log {\it S} curves published by
\citet{giacconi01}.

We initially performed a simulation to investigate the source
statistics in the medium (1--2 keV) band.  Here the comparison was
between the number of actual detections in this band and the numbers
predicted on the basis of the 0.5--2 keV log {\it N} -- log {\it S}
relation\footnote{In the simulation we constrained the flux range from
which input sources were randomly selected, subject to the log {\it N}
-- log {\it S} weighting, to be $9 \times 10^{-16}\ergcms$ to $2
\times 10^{-13}\ergcms$.}.  The simulation involved distributing
sources over a sky area encompassing the full field of view of the NGC
55 observations with a surface density and flux distribution
consistent with a random sampling of the input log {\it N} -- log {\it
S} curve.  The X-ray flux of each simulated source was then converted
to counts by folding in the 1--2 keV exposure maps, applying the
relevant 0.5--2 keV flux to 1--2 keV counts conversion (assuming the
spectral form defined earlier) and finally Poisson deviating the
predicted count.  A given source was deemed to be `detected' if its
estimated count exceeded a threshold value in either the pn or MOS
channels (or both) in either of the EPIC observations. The full
process involved the determination of the average number of source
detections in 5000 simulation runs.  The threshold count used in the
analysis was set by requiring the number of simulated detections to
match the number of actual detections for the off-galaxy regions of
the EPIC fields.  In practice we found that a threshold of 15 counts
gave a reasonable approximation to the detection criteria actually
employed in extracting 1-2 keV band sources.

On the basis of this analysis we conclude that in the medium band, the
number of sources detected within the \d25 ellipse of NGC 55 exceeds
the source count prediction by $17$, that is $\sim 55$ per cent of the
total. However, this must represent a lower limit since the simulation
ignores the effect of absorption in the disc of the galaxy, which will
tend to suppress the count rates of background sources below the
predicted levels.  Allowing for column densities within the disc of
NGC 55 of up to $10^{22} \rm~cm^{-2}$, we estimate that the above
fraction should be increased by a further $5-10$ per cent, implying
that roughly two-thirds of the medium band sources within the \d25
ellipse are probably associated with the galaxy.  These sources
presumably comprise the bright end of the intrinsic X-ray source
population of NGC 55.

The extension of the above analysis to the soft band is complicated by
the increased impact of the absorption in the disc of NGC 55. For the
hard band, similar results to the above were obtained, albeit with
poorer statistics. However, we consider that few of the \d25 sources
detected solely in the soft band are likely to be background objects
given the softness of their spectra, whereas the hard-band only
detections represent just a few sources in total. In summary, we
estimate that there may be 10-15 background interlopers in the sample
of 42 sources detected within the \d25 ellipse of NGC 55.

\begin{figure*}
\begin{center}
\scalebox{0.4}{{\includegraphics[angle=270]{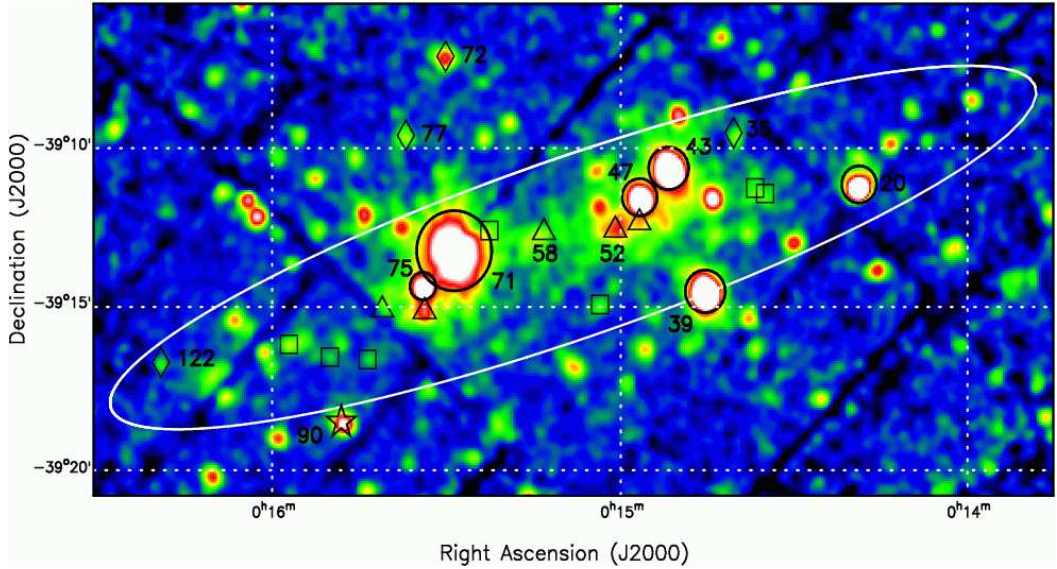}}}
\end{center}
\caption{\label{bsrcs} The central regions of the \xmmn image of the
NGC 55 field in a broad (0.3 -- 6 \,keV) bandpass. The ellipse marks
the optical extent of the galaxy as measured by the \d25 isophote.
The six brightest X-ray sources are highlighted with circles denoting
the spectral extraction radius used. The triangles mark the positions
of sources with hardness ratios commensurate with SNRs, whereas the
squares mark very soft sources (see sec. 5). An X-ray source possibly
located within a globular cluster (\citealt*{beasley00}) lies at the
position of the star symbol. Sources identified with background
galaxies are indicated by diamonds.}
\end{figure*}

\subsection{Cross-correlation with other catalogues}

We have cross-correlated the current source list with other
multi-wavelength data, including the catalogues available in the NED
and SIMBAD databases, by searching for matches within the 3$\sigma$
error radius (including a 1\arcsec systematic error) of each \xmm
source position. The results are summarised in Table~\ref{crosscorr}.
Of the 14 sources with possible identifications, 6 lie within the
confines of the \d25 ellipse of NGC 55.  Four of these are likely to
be objects within the galaxy itself, \ie sources 48, 52 and 58 which
coincide with radio sources and have hardness ratios commensurate with
supernova remnants (see sec. 5), and source 71, a likely black-hole
X-ray binary system in NGC 55 (Paper I). The other two `identified'
\d25 sources are classed optically as galaxies and are presumably
background Active Galactic Nuclei (AGN). Of the remaining 8 identified
sources, 6 are similarly associated with background galaxies, whilst
source 90, which lies just outside the \d25 ellipse, is coincident
with a possible globular cluster located in the halo of NGC 55
(\citealt{liller83}; \citealt{olsen04}) and source 39 is coincident
with a variable (foreground) star. Fig.~\ref{bsrcs} shows the central
part of the 0.3 -- 6 \,keV \xmmn image with sources of interest marked
with different symbols.

Table~\ref{crosscorr} also lists the matches between the \xmmn sources
and previous \rosat X-ray source detections (again within a 3$\sigma$
position error, but also incorporating 15/10\arcsec systematic
uncertainties in the \rosat PSPC/HRI data respectively).  A total of
15 X-ray sources within the \d25 ellipse have previously been
catalogued by {\it ROSAT}, as have 12 outside this region.  Finally,
we performed a brief comparison of these sources to the NGC 55 source
list from the \chandra observation (Obs ID 2255), produced via the
standard \textsc{ciao} processing tools.  From this comparison we
found that our \xmmn sources 29, 34, 55 and 56 are potentially subject
to source confusion, as they could be resolved into two separate
sources by the sub-arcsecond spatial resolution of \chandra.


\begin{table}
\begin{center}
\caption{\label{sigdet} The number of sources detected in each energy band in the pn
and MOS images.}
\begin{tabular}{cllll}
\hline
    &\multicolumn{4}{c}{Energy Band}\\ \cline{2-5}
Camera  &SOFT       &MEDIUM         &HARD       &S/M/H\\
\hline
pn  	&25[91]     &25[93]     &18[55]     &35[119]\\
MOS 	&19[70]     &22[83]     &20[64]     &32[103]\\
pn/MOS  &29[105]    &30[110]    &23[76]     &42[137]\\
\hline
    &           &       &           &\\
\end{tabular}
\begin{minipage}[t]{2.9in}
{\sc Notes:} The numbers quoted are for the \d25 ellipse with the
full-field values given in square brackets.  The total number of
individual sources detected in various instrument and energy band
combinations is also indicated.
\end{minipage}
\end{center}
\end{table}


\begin{table*}
\begin{center}
\caption{\label{crosscorr} Matches of the \xmmn sources with sources in other
catalogues.}
\begin{tabular}{lllcc}
\hline

\xmm ID &\multicolumn{2}{l}{\rosat ID}  	    &Other ID  			&Source type\\ 
        &PSPC           		&HRI        &                   	&\\
\hline
8       &           			&           &APM B001129.07-391943.0    &Galaxy\\
17      &N55-1          		&           &                   	&\\
20*     &R1,S19, N55-2  		&N55-2      &                   	&\\
27*     &N55-3          		&           &                   	&\\
29      &S18            		&           &                   	&\\
35*     &           			&           &BS2000-17         		&Galaxy\\
38*     &R2,S17, N55-4  		&N55-4      &                   	&\\
39      &S16, N55-5     		&N55-5      &UY Scl               	&Variable Star\\
41      &N55-6          		&N55-6      &                   	&\\
42      &S14, RXJ 001451.5-392417	&           &                   	&\\
43*     &R3, S15,N55-7  		&N55-7      &                   	&\\
46      &1WGA J0014.9-3916, N55-8 	&           &                   	&\\
47*     &R4, S13         		&N55-9      &                   	&\\
48*     &                 		&           &HDW86 - RS           	&Radio Source \\
52*     &R5, S11, N55-10      		&           &HDW86 - RS           	&Radio Source\\
55*     &N55-11         		&           &                   	&\\
58*     &           			&           &HK83-08, HDW86 - RS     	&\hii region, Radio Source\\
71*     &R6,S7, N55-14       		&N55-14     &XMMU J001528.9-391319      &ULX, XRB\\
72      &           			&           &LCRS B001300.1-392355      &Galaxy\\
74*     &R7, N55-15      		&           &                   	&\\
77      &           			&           &LCRS B001306.8-392624      &Galaxy\\
79*     &N55-16         		&           &                   	&\\
81      &1WGA J0015.6-3926   		&           &                   	&\\
84*     &N55-17         		&           &                   	&\\
87      &N55-18         		&           &                   	&\\
90      &N55-19         		&           &LA43                   	&Globular cluster\\
94      &           			&           &APM B001321.45-391912.8    &Galaxy\\
101*    &N55-20         		&           &                   	&\\
110     &N55-22         		&           &                   	&\\
112*    &N55-23         		&           &                   	&\\
117*    &N55-24         		&           &                   	&\\
118     &1WGA J0016.1-3920, N55-25	&           &                   	&\\
122*    &           			&           &LCRS B001349.4-393325      &Galaxy\\
123     &           			&           &APM B001352.34-394040.4    &Galaxy\\
125     &1WGA J0016.5-3905   		&           &           		&\\
132     &           			&           &LCRS B001412.2-393501      &Galaxy\\
137     &1WGA J0017.0-3910   		&           &           		&\\
\hline
\end{tabular}
\begin{minipage}[t]{5.9in}
{\sc Notes:} \rosat ID numbers refer to \citet{read97} (R\#),
\citet{schlegel97} (S\#) and \citet{roberts97} (N55-\#). Other
references: APM - \citet{maddox90}; 1WGA - \citet{white00}, BS2000 -
\citet{beasley00}; UY Scl - \citet{perryman97}; RXJ -
\citet*{barber96};
HK83- \citet{hodge83}; HDW86 - \citet{hummel86} (\ie the triple
radio source);
XMMU - \citet{stobbart04}; LCRS - \citet{shectman96}; LA43 - \citet{liller83}.
\end{minipage}
\end{center}
\end{table*}

\section{The brightest discrete X-ray sources}

Six catalogued sources with observed fluxes in excess of $10^{-13}
\ergcms$ lie within or close to the \d25 ellipse of NGC 55.  These
sources (20, 39, 43, 47, 71 and 75 in Table 2) have been detected with
sufficiently high count rates to permit reasonably detailed spectral
and temporal analyses of their X-ray properties.  We presented a study
of the brightest source (71) in Paper I.  Here, we analyse four of the
next five brightest sources, excluding only source 75 for reasons of
practicality.  The emission from this latter source was contaminated
by the wings of the point spread function (PSF) of source 71 in both
observations (and furthermore it lay at the very edge of the field of
view in the second observation, where it was detected only in the MOS
data).  We do include source 20, although it was only within the field
of view in the second observation, and source 39 which as previously
noted may be identified with a foreground star.

\subsection{Light curves}

\begin{figure}
\begin{center}
\scalebox{0.35}{{\includegraphics[angle=270]{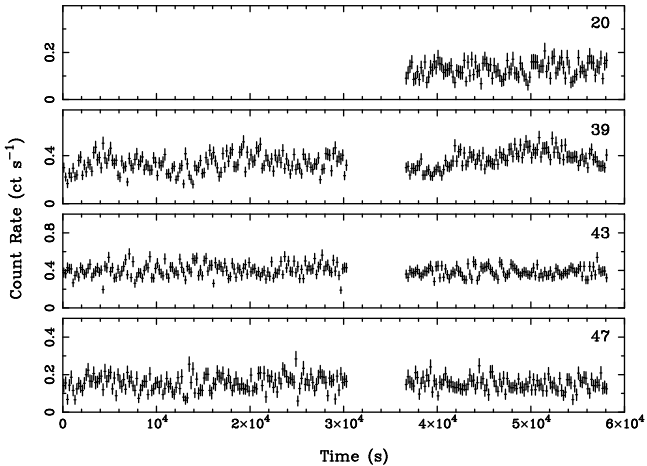}}}
\end{center}
\caption{\label{lcs} The background-subtracted 0.3 -- 6 \,keV
light curves of four bright sources, displayed in 200 s time bins.
The light curves are based on the combined data from the MOS-1,
MOS-2 and pn cameras. The error bars correspond to $\pm 1\sigma$.}
\end{figure}

We extracted background-subtracted light curves for each source based
on the combined 0.3 -- 6 keV data from the three EPIC cameras. For
sources 20 and 47 we used circular source extraction apertures with
radii of 35\arcsec. For the marginally-brighter sources 39 and 43, we
used extraction regions with a 40\arcsec radius.  The background data
were extracted from same-sized apertures, positioned in the nearest
source-free regions visible in all three detectors. We show the
resulting light curves with 200~s time binning in Fig.~\ref{lcs}.  In
each case we correct the count rate from each observation for
vignetting effects so as to plot the equivalent on-axis count rate.

We have investigated the short-term X-ray variability of each source
by carrying out a \chisq~ test (against the hypothesis of a constant
flux) on the raw 200~s binned light curves.  We also performed a
Kolmogorov-Smirnov (K-S) test to search for additional variability
characteristics, such as low amplitude underlying trends in the light
curve, that a \chisq~ test might not detect.  For the latter purpose
we used only the pn light curves with 1s time bins.  The results of
both tests are summarised in Table~\ref{var}.  It appears that source
39 (the foreground star) is the only object in the set for which there
is strong evidence for short-term X-ray variability (greater than 99.9
per cent probability in each test).  Of the three remaining sources,
sources 20 and 47 show limited evidence (at $\ga 2\sigma$
significance) of variability in one of the tests.  However, source 43
shows no evidence at all of short-term variability.

All four sources were also previously detected in both the \rosat PSPC
(performed on 22 -- 24 November 1992) and HRI (12 -- 14 December 1994)
observations indicating that these are all relatively persistent
sources.  A comparison of the measured X-ray fluxes (based on the
simple spectral models derived in the next section) reveals that
sources 20, 39 and 43 varied by factors of $\sim 2$, $\sim 2$ and
$\sim 3$ respectively, but that the flux of source 47 remained
remarkably steady at $1.4 \pm 0.2 \times 10^{-13}
\ergcms$ (0.5 -- 2 keV) over the 9-yr baseline.

\begin{figure}
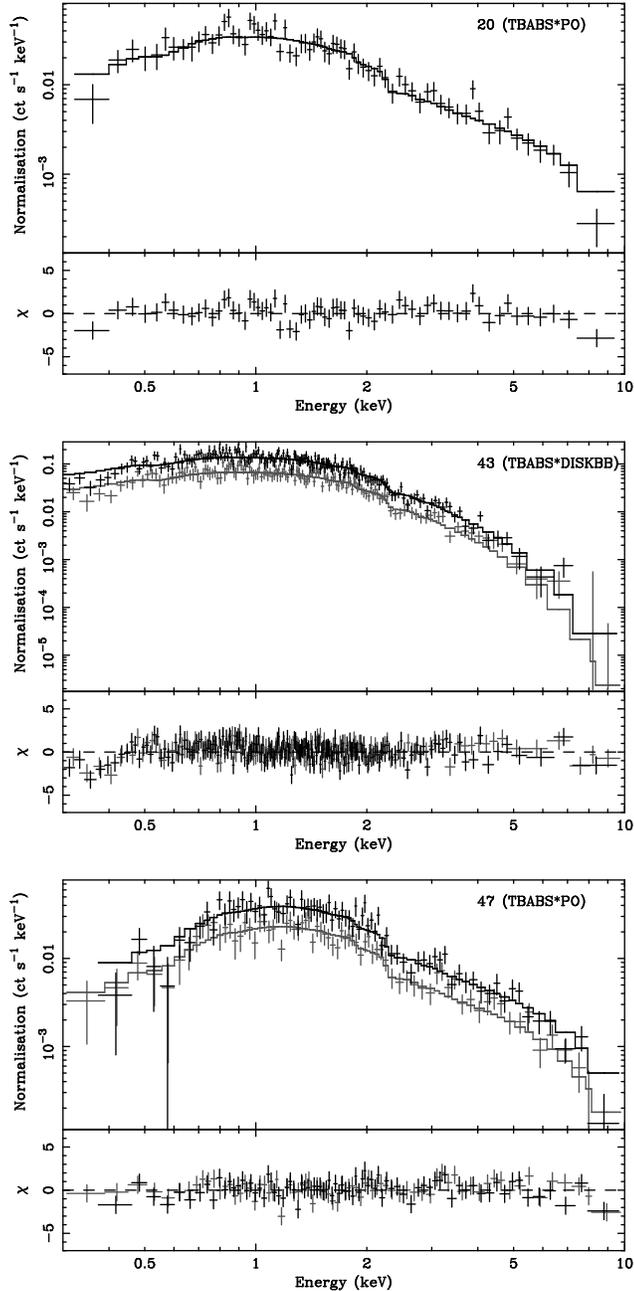

\begin{center}
\scalebox{0.35}{{\includegraphics[angle=270]{figure5a.ps}}}\vspace*{0.3cm}
\scalebox{0.35}{{\includegraphics[angle=270]{figure5b.ps}}}\vspace*{0.3cm}
\scalebox{0.35}{{\includegraphics[angle=270]{figure5c.ps}}}
\end{center}
\caption{EPIC pn count rate spectra and $\Delta\chi$ residuals with respect to the
best-fitting single component model specified in Table~\ref{specres1}, for sources 20,
43 and 47. Data from the first and second observations are shown as grey and black
respectively.}
\label{specs}
\end{figure}

\begin{figure}
\begin{center}
\scalebox{0.35}{{\includegraphics[angle=270]{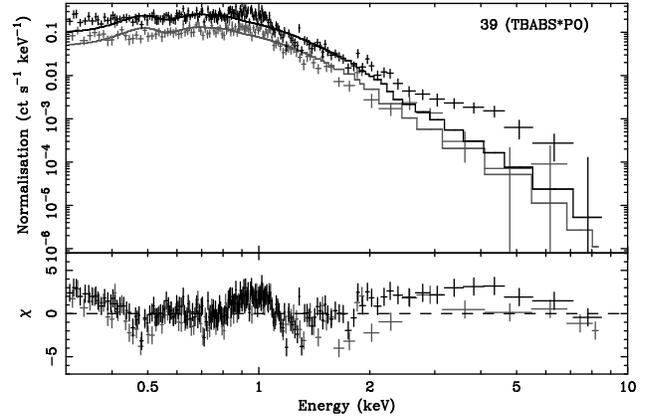}}}
\end{center}
\caption{\label{fgsspec} EPIC pn count rate spectra and $\Delta\chi$ residuals for a
simple power-law continuum model fit to source 39 (the putative
foreground star).  The data are displayed as in Fig.~\ref{specs}.}
\end{figure}

\begin{table}
\begin{center}
\caption{\label{var} Tests for short-term variability}
\begin{tabular}{ccccc}
\hline
Source      &\multicolumn{2}{c}{\chisq statistic}   &K-S statistic\\ 
        &\chisq/dof &P$_{\chi^2}$(var)  &P$_{K-S}$(var)\\
\hline
{\bf Obs 1}\\
\hline
39      &228.5/151  &$>99.9$ per cent   &$>99.9$ per cent\\
43      &155.2/151  & --            & -- \\
47      &153.2/151  & --            &$98$ per cent\\
\hline
{\bf Obs 2}\\
\hline
20      &137.3/107  &$97$ per cent        & -- \\
39      &274.5/107  &$>99.9$ per cent   &$>99.9$ per cent\\
43      &114.7/107  & --            & -- \\
47      &105.8/107  & --            & --\\

\hline
\end{tabular}
\begin{minipage}[t]{3in}
{\sc Notes:}
The probability that a source is variable in either observation,
according to the \chisq and K-S tests.  We only quote probabilities
greater than 95 per cent so as to highlight possible variability.
\end{minipage}
\end{center}
\end{table}

\subsection{X-ray Spectra}

X-ray spectra were extracted in the 0.3 -- 10 \,keV band, using
circular extraction apertures with radii of 35\arcsec (sources 20 and
47) or 40\arcsec (sources 39 and 43) (cf. Fig.~\ref{bsrcs}).
Background data were again extracted from nearby source-free regions,
in this case with radii of 50\arcsec for sources 39 and 43, and
60\arcsec for sources 20 and 47; as far as possible, we chose regions
located at similar distances from the pn readout nodes to the sources,
and free of diffuse emission from the galaxy - see sec. 6.  We used
the {\sc sas} task {\sc especget} to produce a source and background
spectrum together with the Ancillary Response File (ARF) and
Redistribution Matrix File (RMF) required in the spectral fitting in
each case. The spectral analysis was performed using the spectral
fitting package {\sc xspec} (v.11.3.0).  Spectral channels were
grouped so as to give a minimum of 20 counts per bin, thus ensuring
\chisq~ statistics are valid.  The pn, MOS-1 and MOS-2 spectra from
both observations were fitted simultaneously, but we included constant
multiplicative factors in each model to allow for calibration
differences between the cameras.  This value was frozen at unity for
the pn data and allowed to vary for the MOS detectors, with the values
typically agreeing within 15 per cent.

Initial fits to the X-ray spectra of the three bright sources, most
likely associated with NGC 55 (\ie 20, 43, 47), were made using single
component models in {\sc xspec}.  These included power-law ({\sc po}),
blackbody ({\sc bb}), multi-colour disc blackbody ({\sc diskbb}: an
approximation of the optically-thick thermal X-ray spectrum of an
accretion disc [\citealt{mitsuda84}, \citealt{makishima86}]),
bremsstrahlung ({\sc bremss}) and solar-abundance optically-thin
thermal plasma ({\sc mekal}) spectral forms. An intervening absorption
column ({\sc tbabs}, utilising the relative interstellar abundances
and absorption cross-sections tabulated by \citealt{wilms00}) was
applied in each case. Typically, the best fits were obtained with
either the {\sc po} or the {\sc diskbb} models.  These results are
summarised in Table \ref{specres1}, which also lists the observed 0.3
-- 10 keV fluxes and the observed and intrinsic (unabsorbed)
luminosities for the best fitting model. The spectra of sources 20 and
47 are best described by a simple absorbed power-law continuum with
$\Gamma \sim$ 1.7 and 1.8 respectively, although in both cases the
{\sc diskbb} model also provides a statistically acceptable fit (with
$kT_{in} \sim 1.7$ and 1.5 keV respectively).  However, a power-law
model is rejected (at $> 99.99$ per cent probability) for source 43,
which is instead very well fitted by a {\sc diskbb} model with an
inner disc temperature of $\sim 0.8$ keV. We note that a marginally
better fit for source 43 was achieved using an absorbed bremsstrahlung
model (with parameter values $kT \sim 1.9$\,keV, \nh $\sim 2.4 \times
10^{21}$ cm$^{-2}$ and $\chi^2$/dof of 546.5/573).  Fig.~\ref{specs}
shows the spectra and best fitting single component model in each
case.

Given the high apparent X-ray luminosities of these sources, \ie close
to or exceeding the Eddington limit for spherical accretion on to a
$\sim 1.4$ \Msun neutron star, we also attempted to fit the data with
the canonical model for black hole X-ray binary spectra, namely a
combination of {\sc diskbb} and {\sc po} components (subject to
absorption).  This two component model provided an improvement to the
spectral fits for all three sources, albeit with varying degrees of
improvement ($\Delta \chi^2 \sim 4.5 - 44$ for two extra degrees of
freedom).

The resulting inner-disc temperatures for sources 20 and 47 are
unusually low for `normal' luminous stellar-mass black hole binaries
($kT_{in} \sim 0.7 - 2$ keV, {\it cf.} \citealt{mcclintock03}), hence
we also tried alternative ({\sc bb}, {\sc mekal}) models for the soft
component. These produced equally good fits to the data, although we
note that in all cases the soft components only account for a small
fraction ($\sim 6$ per cent) of the total observed X-ray flux.  In the
case of the fits to source 43, we obtained an even more puzzling
result; the power-law fitted to the soft end of the X-ray spectrum.
Similar results have been seen for several ultraluminous X-ray
sources, including source 71 in the same host galaxy
(e.g. \citealt{stobbart04}; \citealt{roberts05}), where the presence
of such a power-law component is difficult to explain physically, and
as such the power-law is most likely a proxy for another soft
component.  We therefore attempted both the alternative model fits as
per sources 20 and 47, that also resulted in improved fits to source
43.

We quantify the statistical significance of the improvement offered by
these additional components (above the simple single component fits)
using the prescription described in sec. 5.2 of
\citet{protassov02}\footnote{The F-test {\bf should not} be used in
the case of additive model components, because it does not follow its
expected theoretical reference distribution in this case
\citep{protassov02}.}. This uses a Monte Carlo recipe for {\sc xspec}
to simulate the expected reference distribution of F-values based on
an input power-law spectrum.  This reference distribution yields a
confidence probability for the fit improvement, as shown in Table
\ref{specres2}. Our simulations show that the two-component modelling
provided significant improvements ($\ga 3 \sigma$, at the best
resolution of our simulations) over a simple power-law model for
source 47, and over a {\sc diskbb} for source 43.  However, the
$\Delta \chi^2 \sim 5$ improvements for source 20, for two extra
degrees of freedom, were not significant (at a probability of $< 90$
per cent), hence we do not show these fits in Table \ref{specres2}.
We discuss the plausibility of these spectral solutions in sec. 7.

Fig.~\ref{fgsspec} shows that the final source (39), does not have a
spectrum with a simple continuum shape.  As this source is coincident
with the $m_v = 11.5$ variable foreground star UY Scl and is
presumably a stellar coronal source, we first attempted to fit its
X-ray spectrum with a thermal plasma ({\sc mekal}) model.  However, a
single temperature {\sc mekal} model resulted in an unacceptable fit,
with a reduced $\chi^{2}$ (i.e. $\chi^{2}_{\nu}) \sim 1.7$, even after
allowing its abundance to vary from solar.  This was due, in part, to
an apparent increase in the hardness of the source at energies $> 2$
keV in the second observation ({\it cf.} Fig.~\ref{fgsspec}).  We
attempted to model this by allowing the {\sc mekal} component
temperature and normalisation to vary between the two observations,
but this approach proved unsuccessful.  However, the addition of a
second, hotter {\sc mekal} component, with a normalisation free to
vary between the observations did lead to a significant improvement in
the fit ($\chi^{2}_{\nu} \sim 1.2$), although residuals remained
evident above 2 keV.  An acceptable fit was finally produced by the
addition of a third {\sc mekal} component, and again allowing the
normalisation of the hottest {\sc mekal} to vary between the
observations (whilst keeping the others constant, and constraining
them to the same abundance).  We detail the best fit in
Table~\ref{star}.  The variation in the hottest component was
considerable, implying at least a factor $10$ increase in its strength
between the two observations, although even at its brightest level (in
the second observation) its contribution to the total 0.3 - 10 keV
source flux amounts to only $\sim 20$ per cent.

The low absorption column (set at Galactic) and multi-temperature
thermal plasma spectrum provide strong support for the origin of the
X-ray emission to be in the stellar corona of an active star in our
own Galaxy.  Temperatures of $\sim 0.3$ and 1 keV are typical for such
objects, however a prominent emission component at temperatures at or
above 2 keV is commonly seen only in young stars or active binaries
(such as RS CVns), and rarely in main sequence stars except during
flares \citep{gudel04}.  A published optical spectrum reveals radial
velocity variations of $\sim 40 \kmsec$ (\citealt{solano97}),
characteristic of a close binary star system, and the 2MASS colours of
UY Scl reveal it to have a late (possibly K-giant) spectral
type. Therefore, a plausible interpretation of the X-ray and optical
properties of the variable star UY Scl is that it is a
previously-unidentified RS CVn system.


\begin{table*}
\begin{center}
\caption{\label{specres1} Spectral fitting of single component models for three bright
sources in NGC 55}
\begin{tabular}{cccccccccccc}
\hline
    	&\multicolumn{3}{c}{{\sc po}$^{a}$}                   		&   &\multicolumn{3}{c}{{\sc  diskbb}$^{a}$}       			&   &$f_X$$^{b}$        	&\multicolumn{2}{c}{$L_X$$^{c}$}\\ \cline{2-4} \cline{6-8} \cline{11-12}
Source  &\nh$^{d}$		&$\Gamma$       	&\chisq /dof    &   &\nh$^{d}$      		&$kT_{in}$      	&\chisq /dof    &   &Obs            		&Obs            &Unabs\\
\hline
20  &2.56$^{+0.44}_{-0.39}$ 	&1.67$^{+0.09}_{-0.08}$ &\bf{125.6/119} &   &0.80$^{+0.26}_{-0.24}$    	&1.65$^{+0.14}_{-0.12}$ &130.9/119    	&   &3.52$^{+0.23}_{-0.27}$ 	&1.33$^{+0.09}_{-0.10}$ &1.67$^{+0.08}_{-0.15}$\\
43  &4.23$\pm 0.20$		&2.70$\pm 0.05$ 	&715.0/573   	&   &1.25$\pm$0.11      	&0.79$\pm$0.02      	&\bf{551.3/573} &   &6.33$^{+0.11}_{-0.14}$ 	&2.40$^{+0.04}_{-0.09}$ &2.90$^{+0.03}_{-0.09}$\\
47  &4.64$^{+0.47}_{-0.44}$ 	&1.84$\pm 0.07$ 	&\bf{252.0/276} &   &2.05$^{+0.28}_{-0.26}$ 	&1.50$^{+0.09}_{-0.08}$ &278.2/276    	&   &4.60$^{+0.20}_{-0.24}$	&1.74$^{+0.08}_{-0.09}$ &2.55$^{+0.05}_{-0.19}$\\
\hline
    &   &                   &               &   &                   &               &           &   &           &   & \\
\end{tabular}
\begin{minipage}[t]{6.2in}
{\sc Notes:} $^{a}$ Spectral models are abbreviated to {\sc xspec}
syntax: {\sc po} --  power-law continuum; {\sc diskbb} --
multicolour disc black body emission; $^{b}$ Observed 0.3--10
\,keV X-ray flux ($\times 10^{-13}$ erg cm$^{-2}$ s$^{-1}$);
$^{c}$ Observed and intrinsic 0.3--10 \,keV X-ray luminosity
($\times 10^{38}$ erg s$^{-1}$); $^{d}$ Column density (including
Galactic, $\times 10^{21}$ cm$^{-2}$). The best fitting model for
each source is highlighted by showing the \chisq /dof values in
bold face.
\end{minipage}
\end{center}
\end{table*}

\begin{table*}
\begin{center}
\caption{\label{specres2} Two-component spectral modelling results}
\begin{tabular}{clcccccccc}
\hline
Source  &Model$^{a}$		&\nh			 &$\Gamma/kT/kT_{in}$$^{b}$	&$kT_{in}/\Gamma$$^{c}$         &\chisq /dof    &$\Delta$\chisq $^{d}$  &\multicolumn{2}{c}{$f_X$ fraction$^{e}$}\\ \cline{9-10}
    	&       		&                   	 &                   		&                       	&           	&               	&Soft       &{\sc po}\\
\hline																				
43  &{\sc po}+{\sc diskbb}     	&5.62$\pm$0.11		 &$4.91^{+0.74}_{-0.78}$	& 0.82$\pm$ 0.04		& 507.3/571	& 44.0			&0.27	    &0.73 \\
    &{\sc bb}+{\sc diskbb}     	&$2.57^{+0.86}_{-0.42}$	 &0.14$\pm$0.02			&$0.81^{+0.03}_{-0.04}$		& 513.5/571	& 37.8			&0.1	    &0.9 \\
    &{\sc mekal}+{\sc diskbb}	&$1.67^{+0.25}_{-0.21}$	 &$0.28^{+0.06}_{-0.04}$	&0.80$\pm$ 0.02			& 522.9/571	& 28.4			&0.04	    &0.96 \\
47  &{\sc diskbb}+{\sc po}     	&10.97$^{+1.56}_{-2.23}$ &0.11$^{+0.02}_{-0.01}$        &2.08$^{+0.09}_{-0.13}$ 	&234.1/274      &17.9               	&0.06       &0.94\\
    &{\sc bb}+{\sc po}          &10.41$^{+2.36}_{-2.21}$ &0.10$\pm$0.01          	&2.06$^{+0.11}_{-0.12}$         &234.1/274      &17.9               	&0.06       &0.94\\
    &{\sc mekal}+{\sc po}       &9.33$^{+2.39}_{-2.52}$  &0.24$^{+0.09}_{-0.05}$ 	&1.99$^{+0.12}_{-0.14}$         &234.3/274  	&17.7               	&0.06       &0.94\\

\hline
    &               &               &                       &               &           &               &           &       &\\
\end{tabular}
\begin{minipage}[t]{6.2in}
{\sc Notes:} $^{a} $ Spectral models are abbreviated to {\sc xspec}
syntax: {\sc po} and {\sc diskbb} as before, {\sc bb} - blackbody
continuum, {\sc mekal} - solar-abundance thermal plasma model.  The
dominant soft component is listed first.  $^b$ Value of characteristic
parameter for soft component.  $^c$ Value of characteristic parameter
for hard component.  $^{d} $\chisq improvement over the best-fitting
single component fit, for two extra degrees of freedom.  $^{e}$
Fraction of the total observed flux in the soft and power-law
components.  The significance probability of the fit improvement over
the single component fit is 100 per cent in each case, based on the
resolution of the Monte Carlo simulations described in the text (200
simulations per source).

\end{minipage}
\end{center}
\end{table*}


\section{X-ray colours}

X-ray colour classification has recently been used by
\citet{prestwich03} and \citet{kilgard05} to provide a statistical
distinction between populations of X-ray binaries (XRBs) and supernova
remnants (SNRs) observed in nearby galaxies.  This scheme, which was
originally applied to \chandra observations, is useful when the X-ray
data are too limited for detailed spectral diagnostics to be
considered on a source by source basis.  In this work, we use an
adaptation of this X-ray colour classification scheme tuned to \xmmn
data \citep{jenkins05}\footnote{An alternative classification scheme
developed specifically for \xmmn data is described in
\citet{pietsch04}.}.

The basic approach is to divide the HR1 versus HR2 colour space into
sub-regions so as to distinguish different classes of X-ray source,
albeit within the limits set by the overlapping spread of spectral
form which characterises the various populations. The measurement
uncertainties on the X-ray colours (HR1 and HR2) also serve to blur
the population boundaries. Here we divide the set of sources detected
within the \d25 ellipse of NGC 55 into four broad categories of
source: `absorbed', `XRB', `SNR' and `other' sources using the same
criteria as employed by \citet{jenkins05} in their study of
M101\footnote{Here `other' sources refers to the background and
indeterminate soft/hard source categories in Table 3 of
\citet{jenkins05}.}). In the event, there were no sources in NGC 55
falling in the `other' category and, for our current purpose, we have
combined the `absorbed' and `XRB' sources into a single `XRB'
category.  

However, we have also added a further sub-division of the soft X-ray
source population using the criterion, HR1$<-0.8$, to classify sources
as `very soft' sources (VSSs)\footnote{This definition is similar,
although not identical, to the definition of very soft sources used by
\citet{distefano03}.  For example, they base their work on \chandra
ACIS-S data, use slightly different bands and have additional
selection criteria.}.  This class could contain a menagerie of exotic
sources, including true supersoft sources (SSSs).  SSSs emit X-rays
predominantly below 0.5 keV (\citealt{greiner00}).  Most are believed
to be accreting white dwarfs undergoing nuclear burning on their
surface (\citealt{vanden92}; \citealt{rappaport94}).  

Other possibilities for very soft emitters include some SNRs, accreting
neutron stars with large photospheres, intermediate-mass black holes,
symbiotic systems, the hot cores of young planetary nebulae and
stripped cores of tidally disrupted stars (\eg \citealt{distefano04}).
Our classification will also contain those
sources classed as `quasi-soft' by \citet{distefano04}, which could
include slightly hotter and/or absorbed variants on the above source
types.

However, while our chosen energy bands provide good signal to noise
coverage for the data in general, extra soft energy sub-bands are
required for more confident identifications of true SSSs.  This has
been explicitly demonstrated for \xmmn data from the galaxies M33
\citep{pietsch04} and M31 \citep{pietsch05a} and the emission of
optical novae within them \citep{pietsch05b}.  We therefore extracted
0.3 - 0.5 keV images from each dataset, and performed aperture
photometry at the positions of the VSSs.  On the basis of this, we
were able to identify sources 53 and 63 as good candidate SSSs due to
the vast majority of their detected source counts originating below
0.5 keV.  The remaining 5 VSSs were all dominated by 0.5 - 1 keV
counts.

On the basis of this X-ray colour classification scheme there are 30
XRBs, 5 SNRs and 7 VSSs within the \d25 ellipse of NGC 55. Earlier we
concluded that 10-15 of the sources within \d25 ellipse might in fact
be background sources (largely AGN) seen through the disc of the
galaxy. Since the absorption in the disc of NGC 55 will preferentially
suppress the soft emission of background objects, we would expect such
background sources to reside predominantly in the XRB (and `absorbed'
source) colour range. Background objects could therefore account for
between 30 per cent - 50 per cent of the sources in the XRB category.
The ratio of `hard' XRBs to `soft' sources (SNRs, VSSs and `other'
sources) observed in NGC 55 (roughly 60 per cent:40 per cent) is very
similar to that seen in M101 \citep{jenkins05}, after (in both cases)
applying a rough correction for background source contamination. Since
these are respectively edge-on and face-on systems, one might conclude
that to zeroth order the inclination of a galaxy does not impact the
relative prominence of the hard and soft X-ray source populations.  Of
course in practice, factors such as the radial distribution and scale
height of the various populations and the depth of the absorption
associated with the galactic disc, combine to make this a complex
problem.

Perhaps one surprising result of this analysis is that VSSs make up 
one sixth of the total source population detected within the
\d25 ellipse of NGC 55, and could constitute more than one in four
sources after the exclusion of background contamination.  We have
investigated whether this is an unusually large fraction by comparison
to published \xmmn and \chandra studies that detect VSSs in other
nearby galaxies.  Firstly, we note that only $\sim$12 per cent of the
\d25 sources of M101 in an \xmmn observation are classified as
VSSs using our criterion (\cf \citealt{jenkins05}).  As a wider
comparison, we also looked at the VSS population of M101 in a much
deeper \chandra (ACIS-S) observation, as well as those of M83 (face-on
spiral), M51 (interacting galaxy) and NGC 4697 (elliptical)
\citep{distefano04}.  We converted our \xmmn HR1 VSS criterion to the
equivalent \chandra ACIS-S value (via {\sc webpimms}), using the
appropriate energy bands, to facilitate comparison.  Using only our
criterion, we found that the fraction of VSSs were 31 per cent for
M101, 25 per cent for M83, 15 per cent for M51 and 4 per cent for NGC
4697.  At first glance, it therefore appears that there are quite
large variations in the VSS population fractions of different
galaxies.  However, this fraction could be extremely susceptible to
observational effects such as the effective (foreground and intrinsic)
extinction and the luminosity threshold of the observations.  For
example, as typical SSSs in particular have luminosities below $\sim
10^{37} \ergsec$, this latter effect should result in more SSSs being
observed in deeper, more sensitive observations, as may be the case
when comparing the deep \chandra observation of M101 with the
shallower \xmmn data.  The Galactic foreground absorption is certainly
a very important factor on the number of VSSs detected -- this is
small in M101, M83 and NGC 55 but is significantly higher in M51
and NGC 4697.  An additional influence on the number of VSSs detected
is the choice of filter used in the observations.  The \xmmn
observations of M101 were made using the medium filter while the NGC
55 observations utilised the thin filter.  We conclude that although
the differences between galaxies appear large, selection effects may
dominate.


\begin{table}
\begin{center}
\caption{\label{star} Three-temperature {\sc mekal} spectral model for UY Scl.}
\begin{tabular}{lccccc}
\hline
Parameter   				&Value \\
\hline
\nh ($\times 10^{21}$ cm$^{-2}$)    	& 0.155 ({\it fixed\/}) \\
$kT_1$ (keV)                		& 0.33$\pm$0.02 \\
$kT_2$ (keV)                		& 1.00$\pm$0.04 \\
$kT_3$ (keV)                		& 3.44$^{+2.39}_{-1.27}$ \\
Abundance (solar units)         	& 0.43$^{+0.08}_{-0.07}$ \\
\chisq /dof             		& 512.6/463 \\
\hline
\end{tabular}
\end{center}
\end{table}


\section{Residual disc emission}

The individual \xmmn observations were flat-fielded by subtracting a
constant particle rate from the image (estimated by taking an average
of the count rates in the edge regions of the detector not exposed to
the sky) and then dividing by the appropriate exposure map.  In the
same process, bad pixels, a hot column (for the pn detector) and
spurious data along chip gaps were excised. The flat-fielded data from
the two observations were then mosaiced into a single
exposure-corrected image. Here we focus our attention on the pn
soft-band data.


\begin{figure}
\begin{center}
\scalebox{0.305}{{\includegraphics[angle=270]{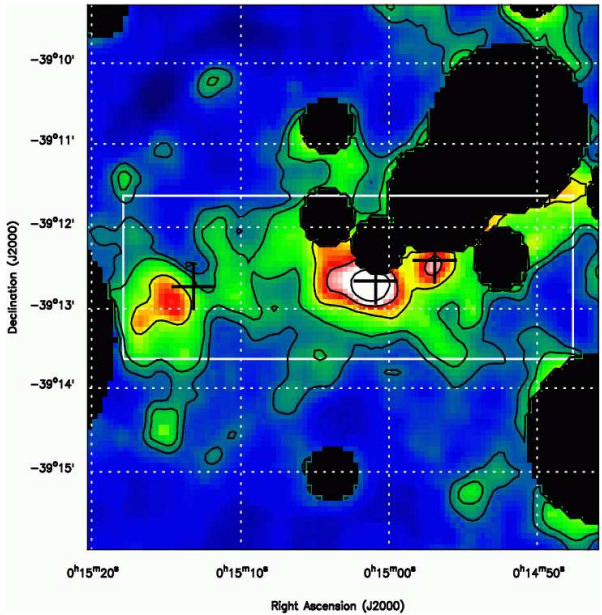}}}
\end{center}
\caption{\label{diffuse} Close up of the composite flat-fielded pn
image in the soft (0.3--1.0\,keV) bandpass. Regions around
catalogued sources have been blanked out so as to reduce the
contamination of the diffuse signal by relatively bright resolved
point sources. An exception was made for three sources in the
region (sources 48, 52 and 58) whose positions are indicated by
crosses, which have HR1 values characteristic of SNRs (see text).
The large rectangle, excluding the blanked out sections,
corresponds to the region used to estimate the diffuse luminosity
of the disc of NGC 55 and from which the diffuse spectrum was
extracted.}
\end{figure}


Fig.~\ref{diffuse} shows the mosaiced pn soft-band image of the
central region of the galaxy. In order to highlight unresolved or
possible diffuse emission (which we refer to hereafter as the residual
disc emission), we have blanked out regions around the catalogued
discrete sources listed in Table \ref{allsrcs}. For this purpose,
circular exclusion regions were used of radius 20\arcsec, 40\arcsec,
1\arcmin or 2\arcmin depending on the strength of the
source\footnote{These sizes were determined empirically based on the
need to suppress high levels of contamination from point sources,
consistent with the size and shape of the point-spread function over
the wide range of offset angles sampled.}. An exception was made for
sources 48, 52 and 58, which have X-ray colours characteristic of SNRs
and which we here consider as part of the diffuse component of the
galaxy. The residual emission is concentrated in the central disc of
NGC 55 encompassing both the bar region and the region to the
immediate east of the bar.  The presence of several relatively bright
X-ray sources makes it difficult to delineate the full extent of this
component along the disc, although on the basis of Fig.~\ref{bsrcs} it
appears to be $^{<}_{\sim} 12$\arcmin. The residual emission is most
evident within $\pm$1\arcmin of the plane, which is well within the
extent of the thin disc component (\citealt{Davidge05}).

We estimate the {\it observed} (0.3--1 keV) luminosity of the disc
within the 6\arcmin $\times$ 2\arcmin rectangular region illustrated
in Fig.~\ref{diffuse} (excluding the blanked out regions) to be $2.5
\times 10^{37} \ergsec$ (based on the observed count rate in 
this band, and energy conversion factors derived from the spectral
analysis discussed below).  Several regions of enhanced surface brightness are
evident within this region, two of which are clearly related to
sources 48 and 52. The easternmost bright region appears to be offset
from source 58 but we note that this source was detected at relatively
low significance only in the MOS soft band channel. Judging from the
pn soft band image, the position of peak X-ray surface brightness is
located approximately 1 s (of time) east of the catalogued position of
source 58 \footnote{The fact that this region was not detected as a
discrete source in the pn data may reflect a limitation in the
background modelling in the vicinity of the brightest X-ray source in
NGC 55.}. The three X-ray bright regions identified above, all
coincide with radio sources and, in fact, X-ray sources 52 and 48
correspond to the core and eastern component of the triple source
identified by \citet{hummel86}. These regions are also prominent in
H$_{\alpha}$ (\citealt{ferguson96}) and in the {\it Spitzer}
far-infrared (24 $\mu \rm m$) image of NGC 55
(\citealt{engelbracht04}) and correspond to sites of current star
formation in the disc of NGC 55. Together the three bright regions
contribute roughly $30$ per cent of the inferred residual disc X-ray
luminosity.

We have extracted spectra representative of the residual disc emission
of NGC 55 from the rectangular region shown in Fig. \ref{diffuse}
(excluding the blanked out regions) using a background spectrum
obtained from a nearby region of the same dimension (from which
catalogued sources were also excluded).  The diffuse spectra were
grouped to give a minimum of 30 counts per bin. The six available
spectral datasets (pn, MOS-1 and MOS-2 data from two observations)
were fitted simultaneously, but with the relative normalisations
untied so as to allow for calibration differences between the cameras.
The spectral analysis was confined to the 0.3--6 keV energy range for
which there was a reasonable signal to noise ratio.

\begin{table*}
\begin{center}
\caption{\label{diffuse_spec} Spectral modelling results for the
residual disc emission}
\begin{tabular}{lccccccccc}
\hline
Model 		  &\nh$^{a}$           &$kT_1$$^{b}$		&$kT_2$$^{b}$		&Z ($\rm Z_{\odot}$)$^{c}$  	&$\Gamma$ 		&\chisq /dof 	&$f_X$$^{d}$    &\multicolumn{2}{c}{$L_X$$^{e}$}\\
        	  &           	       &           		&           		&               		&           		&           	&Obs    	&Obs &Unabs\\
\hline
{\sc mekal}       &2.0$^{+0.4}_{-0.6}$ &0.52$^{+0.09}_{-0.05}$ 	&-- 			&0.04$\pm$0.02  		&--  			&307/217  	&0.94  		&0.36 &0.83 \\
{\sc mekal+po}    &5.1$^{+1.8}_{-2.0}$ &0.21$^{+0.01}_{-0.03}$ 	&-- 			&0.07$^{+0.07}_{-0.03}$ 	&1.9$^{+0.4}_{-0.3}$ 	&230/215  	&1.55 		&0.59 &3.87\\
{\sc mekal+mekal} &5.3$^{+0.9}_{-2.0}$ &0.20$^{+0.06}_{-0.04}$ 	&4.52$^{+9.19}_{-2.12}$ &0.06$^{+0.10}_{-0.03}$ 	&-- 			&231/215 	&1.44 		&0.55 &4.52\\
\hline
            &                           &                       &                       &                       	&          		&               &               &     &\\
\end{tabular}
\begin{minipage}[t]{6in}
{\sc Notes:}
$^{a}$ Column density (including Galactic, $\times 10^{21}$ cm$^{-2}$).
$^{b}$ Temperatures of the cool ($kT_1$) and hot ($kT_2$) thermal components in keV.
$^{c}$ Relative metal abundance of the thermal plasma component.
$^{d}$ Observed 0.3--6 \,keV X-ray flux ($\times 10^{-13}$ erg cm$^{-2}$ s$^{-1}$).
$^{e}$ Observed and intrinsic (absorption-corrected) 0.3--6 \,keV X-ray luminosity ($\times 10^{38}$ erg s$^{-1}$).
\end{minipage}
\end{center}
\end{table*}

\begin{figure}
\begin{center}
\scalebox{0.35}{{\includegraphics[angle=270]{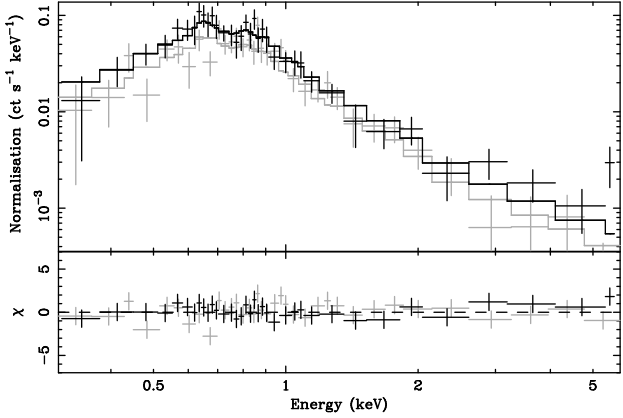}}}
\end{center}
\caption{\label{specdiff} EPIC pn count rate spectra for the
residual disc emission in NGC 55 from the first (grey) and second
(black) observations.  The best-fit {\sc mekal} plus power-law model
is shown along with the corresponding $\Delta\chi$ residuals (lower
panel). NB The data have been rebinned for illustrative purposes using
SETPLOT REBIN in {\sc xspec}}
\end{figure}

Initially, we modelled the X-ray spectra with an absorbed {\sc mekal}
component, with the abundance free to vary. This yielded \nh $\sim 2
\times 10^{21}$ cm$^{-2}$ and $kT \sim 0.5$ keV, with a very low
inferred abundance of $\rm Z \sim$ 0.04 $\rm Z_{\odot}$
($\chi^2$/dof=307/217). An improved fit to the data was achieved with
the inclusion of an additional harder spectral component, either in
the form of a power-law or higher temperature {\sc mekal} emission.
Table~\ref{diffuse_spec} details the best-fit parameters for these
two-component models. The observed pn spectra and the best-fitting
{\sc mekal} plus power-law model is shown in Fig~\ref{specdiff}.

The derived flux using the {\sc mekal} plus power-law model is $\sim
1.6 \times 10^{-13} \ergcms$ (0.3--6 keV), corresponding to an
absorbed luminosity of $6 \times 10^{37} \ergsec$. Roughly half of
this luminosity is contributed by the soft thermal emission. 

The measured column density of $5.1 \times 10^{21}$ cm$^{-2}$ is much
greater than the foreground column in our own Galaxy, and is
consistent with the location of this residual component within the
central disc region of NGC 55. If we correct for this absorption the
inferred luminosity of the soft component increases to $\sim 3 \times
10^{38} \ergsec$. A reasonable hypothesis is that the soft emission
represents a truly diffuse component energised by processes linked to
regions of recent star formation (collision of stellar winds in dense
environments, supernovae and shock heating in SNRs), whereas the
harder spectral component (most appropriately modelled as a power-law)
arises largely in unresolved XRBs.

The derived temperature of the soft residual disc emission in NGC 55
is $\sim 0.2$ keV (Table~\ref{diffuse_spec}).  This is fairly typical
of the softest emission seen in more luminous star-forming galaxies
where the temperature range for diffuse disc components may extend up
to $\sim 2$ keV (\eg \citealt{pietsch01}, \citealt{fabbiano03}). By
analogy to these systems, it is very likely that the hot diffuse gas
in NGC 55 is, in fact, a multi-temperature medium. The very low
abundance derived for this hot gas component
(Table~\ref{diffuse_spec}) is almost certainly an artefact of fitting
a complex spectrum with a very simplified model.

On the one hand the above determination of the luminosity of the
residual disc component in NGC 55 represents only a lower limit
estimate given that we have not measured the true extent of the disc
component and, even within the inner 6\arcmin $\times$ 2\arcmin region
(corresponding to $\rm 3~kpc \times 1~kpc$ at the distant of NGC 55)
we have of necessity excluded regions around bright sources.  On the
other hand the {\it observed} X-ray flux of the residual disc emission
is only $\sim 3$ per cent of that of the resolved sources in the broad
0.3--6 keV band or a $6$ per cent fraction if we consider only the
soft band. Since the point spread function of the \xmmn optics has
wings extending well beyond the blanked-out regions in
Fig.~\ref{diffuse}, some contamination of the residual disc component
by the nearby bright sources is inevitable. However, since the
morphology of the extended emission (in Fig.~\ref{diffuse}) does not
closely track the distribution of the bright sources, we can be
confident that the residual signal (net of the local background) is
not dominated by such contamination, at least in the soft
band. Unfortunately the surface brightness of the residual emission is
too low to apply a similiar imaging test above 1 keV.

Using the parameters from the {\sc mekal} plus power-law model, we
have derived the physical properties of the soft residual disc
component as summarised in Table~\ref{diffuse_params} (this Table
includes a description of the parameters of the diffuse emission
introduced below).  Similar values have been obtained by
\citet{summers03} for the cool diffuse disc component seen in the
Magellanic irregular NGC 4449.  These authors note that the diffuse
medium is likely to be clumpy with a filling factor $f<1$, implying
that the quoted values for $n_{e}$ and $P$ are underestimates, whereas
$M$, $E_{th}$ and $t_{cool}$ are overestimated. The derived pressure
of the hot gas in NGC 55, $P/k \sim 1.5 \times 10^{5} \rm~K~cm^{-3}$,
which is similar to the pressure inferred for the interior of the Loop
1 superbubble within our own Galaxy \citep{willingale03}. This is
broadly consistent with the view that there has been sufficient recent
star formation in the disc of this low-mass dwarf system to form
expanding hot bubbles, which result in the ejection of material out of
the disc of the galaxy into the halo. However, the absence of an
extended extra-planar soft X-ray component in NGC 55 (contrary to the
findings of \citealt{oshima02}) suggests that the gas in such bubbles
cools relatively quickly through adiabatic losses (given the
relatively long radiative cooling timescale in
Table~\ref{diffuse_params}), retaining insufficient energy to power a
superwind of the form frequently seen in systems with star-formation
rates, $\rm SFR > 1 M_{\odot} \rm~yr^{-1}$ \citep{strickland04}.

\begin{table}
\begin{center}
\caption{\label{diffuse_params}Physical parameters of the soft
residual disc component.}
\begin{tabular}{ll}
\hline
Physical property                   	&Value \\
\hline
Temperature, $T$                    	&$2.4 \times 10^6$ K \\
Intrinsic $L_X$                     	&$3.3 \times 10^{38} \ergsec$ \\
Electron density, $n_e$           	&0.029 cm$^{-3}$ \\
Thermal energy, $E_{th}$          	&$6.8 \times 10^{54}$ erg  \\
Mass of hot gas, $M$                	&$5.7 \times 10^6$ \Msun   \\
Pressure, $P$                   	&$2.0 \times 10^{-11}$ dyn cm$^{-3}$   \\
Cooling time, $t_{cool}$            	&$6.5 \times 10^{8}$ yr\\
\hline
\end{tabular}
\begin{minipage}[t]{3in}
{\sc Assumptions:}
$V = 2 \times 10^{65}$ cm$^3$ (cylindrical volume subtending 6\arcmin $\times$ 2\arcmin on the sky); $D = 1.78$ Mpc;
filling factor, $f = 1$; $n_e = (EI/Vf)^{1/2}$, where $EI$ is the emission integral
($norm \times 4 \pi D^2)/10^{-14}$, and $norm$ is the {\sc mekal} normalisation obtained from the
spectral fitting; $ E_{th} = 3 n_e kT V$; $M = n_e m_P V f$; $P = 2 n_e kT$ and
$t_{cool} \sim (3kT)/(\Lambda n_e)$, where $\Lambda = L_X/EI$.
\end{minipage}
\end{center}
\end{table}

\section{Discussion}

\subsection{The brightest sources in NGC 55}

What is the nature of the luminous X-ray source population observed in
NGC 55? This question is most easily addressed for the brightest
subset of X-ray sources, wherein the information provided by X-ray
diagnostics is at its richest. The brightest X-ray source in NGC 55
(XMMU J001528.8-391318, source 71) has an X-ray luminosity $L_X \sim
10^{39} \ergsec$ placing it on the boundary of the ULX classification.
As discussed in Paper I this source is a candidate luminous black hole
X-ray binary.  In sec. 4 we reported the details of four out of the
five next brightest X-ray sources in the NGC 55 field. One of these
sources (39) is spatially coincident with a bright variable foreground
star at the edge of the \d25 ellipse, and its X-ray properties suggest
that it may be a previously unidentified RS CVn system. Of the other
three sources, two are located in the bar region and one in the
western limb of the NGC 55 disc. A combination of their high
luminosities ($L_X > 10^{38} \ergsec$), measured absorption columns
far in excess of the Galactic foreground, and power-law continuum or
accretion disc dominated X-ray spectral shapes (see sec. 4.2), argues
that these sources are likely to be X-ray binaries in NGC 55, though a
clearer detection of short-term variability is required to confirm
this.  Indeed, since the apparent high intrinsic X-ray luminosities of
these sources are close to or in excess of the Eddington limit for a
1.4-M$_{\odot}$ neutron star, the primary accreting object in these
systems may be a black hole. This hypothesis is supported by the X-ray
spectroscopy. We would expect a bright neutron star X-ray binary
spectrum to be described by a combination of a $\sim 1.5$ keV
accretion disc and a $\sim 2$ keV blackbody spectral component with
the latter originating in the neutron star envelope (\cf
\citealt{jenkins04}). Instead our X-ray spectra appear
to be more typical of `low' (hard power-law dominated) and `high'
(thermal accretion disc dominated) black hole spectral states ({\it
cf.} \citealt{mcclintock03}). The long-term ($\sim 9$ year)
persistence of these sources, and their locations close to the
star-forming regions in the disc of NGC 55, suggest that these may be
systems containing a high-mass secondary star.

One interesting result from the spectral analysis of sources 43 and 47
is the detection of a very soft component in each spectrum (a similar
component is seen in source 20, but does not provide a statistically
robust improvement to the fit).  The temperature of this component in
source 47, when modelled by a {\sc diskbb}, is $\sim 0.1$ keV. This is
remarkably similar to the $\sim 0.1 - 0.3$ keV temperatures derived
for the very soft components seen in some nearby ULXs which, it has
been argued, may be evidence for the presence of a $\sim 1000$
M$_{\odot}$ `intermediate-mass' black hole (IMBH)
(e.g. \citealt{miller03}). So should we interpret the soft spectra of
this source (which is one to two orders of magnitude less luminous
than ULXs) in the same way?  If ULXs are accreting IMBHs, then one
might expect to see a range of accretion rates, and hence
luminosities, across the population, as is commonly the case for
Galactic black hole binaries. However, on this basis one might predict
a change in the spectrum to a `low' state as the luminosity decreases,
whereas the spectra for the NGC 55 sources appear very similar to the
apparent `high state' spectrum of IMBH candidates (though in these
cases, as in some ULXs, the power-law slope is anomalously hard for a
true high state; see \citealt{roberts05}). The overall spectral shape
($\Gamma \sim 2$ power-law, $kT_{in} \sim 0.1$ keV {\sc diskbb}) might
actually also describe the `low' state for an ordinary black hole
X-ray binary, which (at a luminosity of $ \sim 1- 2 \times 10^{38}
\ergsec$) must be a plausible alternative for this source.

The shape of the spectrum of source 43 potentially mirrors another
result to emerge from recent ULX studies, an X-ray spectrum described
by a hot $0.7 - 2$ keV {\sc diskbb} with a very soft power-law
dominant at low energies.  The physical implausibility of this
spectral description is discussed by various authors, who note
alternative empirical solutions (e.g. replacing the power-law with a
black-body emission spectrum) or physical models (e.g optically-thick
Comptonisation) that can explain the spectrum (see \citealt{fk05},
\citealt{stobbart06}).

However, a more mundane alternative for the nature of the soft
emission in both sources is suggested by the fact that they are also
well-fitted by a {\sc mekal} component. The derived {\sc mekal}
temperature ($\sim 0.2 - 0.3$ keV in each case) is similar to the
temperature of the residual emission detected in the galaxy.  Also,
the surface brightness required to produce the observed signal within
the source extraction regions is within a factor of $\sim 2$ of the
average surface brightness seen elsewhere in the NGC 55 disc (see
sec. 6). It seems quite likely, therefore, that this very soft
component is simply the underlying emission of the disc of NGC 55
contaminating the source signal\footnote{We note that similar results
are emerging in other \xmmn datasets, \eg the soft excesses of ULXs in
M51 are readily explained as local diffuse emission
\citep{dewangan05}.}.


\begin{table}
\begin{center}
\caption{\label{lxcomp} The components of the integrated X-ray luminosity of NGC 55.}
\begin{tabular}{lc}
\hline
Component                &   $L_X$ (0.3--6 keV) \\
\hline
Brightest (ULX) source   &   $9.4  \times 10^{38} \ergsec$ \\
Next 20 brightest sources&   $6.9  \times 10^{38} \ergsec$ \\
VSSs           		 &   $0.1  \times 10^{38} \ergsec$ \\
Residual soft disc$^{a}$ &   $3.3  \times 10^{38} \ergsec$ \\
Residual hard disc$^{a}$ &   $0.5  \times 10^{38} \ergsec$ \\
\hline
\end{tabular}
\begin{minipage}[t]{3in}
$^{a}$  Corresponding to the 6\arcmin $\times$ 2\arcmin inner disc region.
Corrected for absorption intrinsic to NGC 55 \\
\end{minipage}
\end{center}
\end{table}

\subsection{NGC 55 - a typical Magellanic-type galaxy}

Although the near edge-on orientation has complicated efforts to study
the morphology of NGC 55, the current consensus is that this galaxy is
a dwarf irregular, structurally similiar to the Large Magellanic Cloud
(LMC; see \citealt{Davidge05}).  In order to address the question of
whether NGC 55, from an X-ray perspective, is a typical
Magellanic-type galaxy, we need to consider its overall X-ray
properties.  Earlier we concluded that there might be up to $\sim 20$
XRBs associated with NGC 55 in the sample of sources detected within
its \d25 ellipse, down to an effective luminosity threshold of $2
\times  10^{36} \ergsec$  (0.3 -- 6  keV).  In addition, 7
VSSs were identified in the \d25 sample.  The integrated X-ray
luminosity of these source categories together with the luminosity
inferred for the residual disc emission (including SNRs) is summarised
in Table~\ref{lxcomp}.  (Note that the discrete source luminosities
are from Table~\ref{allsrcs} and are, in effect, observed
luminosities, \ie uncorrected for absorption intrinsic to NGC 55.)

Other nearby  examples of actively star-forming  Magellanic-type
systems include the LMC and NGC 4449, both of which have been
extensively studied in X-rays (\eg \citealt{wang91};
\citealt{vogler97};  \citealt{bomans97};  \citealt{sasaki02};
\citealt{summers03}).  Table~\ref{magellanic}  compares   the
X-ray and other properties of these two galaxies with those of NGC 55
and illustrates the striking similarities of these systems.  Total
mass, mass in neutral hydrogen, star-formation rate (SFR) and the
X-ray luminosity in discrete sources and diffuse components (more
precisely the residual disc in the case of NGC 55) differ by no more
than a factor of $3$ across the three galaxies.  All three systems
would appear to follow the paradigm of strong spatial association of
the bright X-ray emitters with the sites of current star-formation,
with a scaling of X-ray luminosity to SFR broadly consistent with the
$L_{X}$ versus SFR correlation established for much more active
systems (\citealt{ranalli03}; \citealt{gilfanov04}).


\begin{table}
\begin{center}
\caption{\label{magellanic} A comparison of the properties of NGC 55 with 
those of two other Magellanic dwarf galaxies.}
\begin{tabular}{lccc}
\hline
Property                			&LMC 	&NGC 55 &NGC 4449 \\
\hline
Assumed Distant (Mpc)   			&0.05 	&1.78 	&2.93 \\
$L_X$ (sources)$^{a}$ ($10^{38} \ergsec$) 	&5 	&17 	&14 \\
$L_X$ (diffuse)$^{a}$ ($10^{38} \ergsec$) 	&3 	&6 	&10 \\
$M_{25}$$^{b}$  ($10^{9} \rm~M_{\odot}$) 	&3.3 	&8.5 	&4.2 \\
$M_{HI}$$^{c}$  ($10^{9} \rm~M_{\odot}$) 	&0.6 	&2.0 	&1.6 \\
SFR$^{d}$ ($ \rm~M_{\odot}~yr^{-1}$) 		&0.25 	&0.22 	&0.2 \\
\hline
\end{tabular}
\begin{minipage}[t]{3in}
{\sc Notes:} \\
$^{a}$X-ray luminosities from \citet{wang91}; this paper; \citet{summers03} 
respectively \\
$^{b}$ `Indicative' mass of the galaxy within its \d25 diameter
from \citet{kara04}.\\
$^{c}$ HI mass of the galaxy from \citet{kara04}). \\
$^{d}$ Star formation rates from \citet{grimm03}; \citet{thronson87}; 
\citet{engelbracht04} respectively. \\
Quoted values are scaled to the assumed distant. \\
\end{minipage}
\end{center}
\end{table}


\section{Summary}

We have used recent \xmmn EPIC observations to investigate the X-ray
properties of the Magellanic type galaxy NGC 55.  A total of 137 X-ray
sources were detected in the NGC 55 field of which 42 were located
within the optical confines of the galaxy as defined by the \d25
ellipse. The source detections cover a flux range of $\sim 5 \times
10^{-15} - 2 \times 10^{-12} \ergcms$ (0.3-6 keV) corresponding to a
luminosity range of $\sim 3 \times 10^{35} - 9 \times 10^{38} \ergsec$
for sources at the distance of NGC 55.  After allowing for some
contamination of the sample by background AGN, our best estimate,
based largely on X-ray colour classification, is that within the
\d25 region, we detect $\sim 20$ XRBs,  5 SNRs and 7 VSSs
associated with NGC 55. Just outside the \d25 ellipse, we also
detected an X-ray source coincident with a globular cluster in NGC 55.

We have performed detailed X-ray spectral and timing analysis on four
bright X-ray sources ($>500$ counts) in the field of view
(excluding the brightest source which was studied in Paper I). One
of these objects is identified with a Galactic foreground star (UY
Scl) and exhibits many of the properties of an RS CVn system. The
other three are persistent X-ray sources with spectra indicative
of accreting XRBs.

A residual emission component is also evident in the disc of NGC
55, concentrated on the bar region. The extent of this emission is
at least 6\arcmin ( 3 kpc) along the plane of the galaxy and $\pm$1\arcmin
($\pm 500$ pc) perpendicular to the plane. We interpret the soft
component as thermal diffuse emission with  $kT \sim 0.2$
keV associated with regions of current star formation in the disc
of NGC 55.  After correcting for absorption, the X-ray luminosity
of this soft emission is $\sim 4 \times 10^{38} \ergsec$ (0.3--6
keV).

NGC 55 is categorised as a Magellanic-type dwarf galaxy.  From a 
comparison of its properties with those of other nearby Magellanic systems, 
specifically the LMC and NGC 4449, we conclude that, from a high energy 
perspective, NGC 55 is quite typical of its class.


\section*{Acknowledgments}

This work is based on observations obtained with {\it XMM-Newton}, an
ESA Science Mission with instruments and contributions directly funded
by ESA member states and the USA (National Aeronautics and Space
Administration).  The other X-ray data used in this work were obtained
from the Leicester Database and Archive Service (LEDAS) at the
Department of Physics and Astronomy, University of Leicester. The
second Digitized Sky Survey was produced by the Space Telescope
Science Institute, under NASA Contract No. NAS 5-26555. This research
has also used the NASA/IPAC Extragalactic Database (NED) operated by
the Jet Propulsion Laboratory, California Institute of Technology,
under contract with NASA, and the SIMBAD database operated at CDS,
Strasbourg, France. The authors would also like to thank Kevin Briggs
for helpful advice regarding source detection techniques and stellar
X-ray emission. AMS and TPR gratefully acknowledge funding from PPARC.
Finally we thank the referee, Dr Wolfgang Pietsch, for providing a
thorough and informative review of our original manuscript.

\label{lastpage}


{}

\end{document}